%% file: main.tex
\documentclass[sigconf]{acmart}
\AtBeginDocument{%
  \providecommand\BibTeX{{%
    \normalfont B\kern-0.5em{\scshape i\kern-0.25em b}\kern-0.8em\TeX}}}
    
\newif\iffinal
\finaltrue

\usepackage[T1]{fontenc} 
\usepackage[linesnumbered,tworuled]{algorithm2e}
\usepackage{mathtools}
\usepackage{amsmath}
\usepackage{graphicx}
\usepackage{listings}
\usepackage{xspace}
\usepackage{tabularx}
\usepackage{amsthm}
\usepackage{subcaption}
\usepackage{mdframed}
\usepackage[normalem]{ulem}
\usepackage{microtype}
\microtypecontext{spacing=nonfrench}

\usepackage{booktabs}
\usepackage{xcolor}

\usepackage{enumitem}
\usepackage{pdfpages}
\usepackage{url}
\usepackage{color}

\usepackage{tablefootnote}

\newmdtheoremenv[
    innertopmargin=-2pt,
    roundcorner= 10 pt,
]{obs}{Observation}

\input{macros.tex}

\copyrightyear{2020}
\acmYear{2020}
\setcopyright{acmcopyright}
\acmConference[ASIA CCS '20] {15th ACM Asia Conference on Computer and Communications Security}{October 5--9, 2020}{Taipei, Taiwan}
\acmBooktitle{15th ACM Asia Conference on Computer and Communications Security (ASIA CCS'20), October 5--9, 2020, Taipei, Taiwan}
\acmPrice{15.00}
\acmDOI{10.1145/3320269.3384745}
\acmISBN{978-1-4503-6750-9/20/10}

\begin{document}
\fancyhead{}

\title{Hidden in Plain Sight:\\ Obfuscated Strings Threatening Your Privacy}
%
%

\author{Leonid Glanz}
\email{glanz@cs.tu-darmstadt.de}
\affiliation{Technical University of Darmstadt}
\author{Patrick M\"uller}
\email{mueller@cs.tu-darmstadt.de}
\affiliation{Technical University of Darmstadt}
\author{Lars Baumg\"artner}
\email{baumgaertner@cs.tu-darmstadt.de}
\affiliation{Technical University of Darmstadt}
\author{Michael Reif}
\email{reif@cs.tu-darmstadt.de}
\affiliation{Technical University of Darmstadt}
\author{Sven Amann}
\email{research@sven-amann.de}
\affiliation{CQSE GmbH}
\author{Pauline Anthonysamy}
\email{anthonysp@google.com}
\affiliation{Google Inc.}
\author{Mira Mezini}
\email{mezini@cs.tu-darmstadt.de}
\affiliation{Technical University of Darmstadt}

%

%


\begin{abstract}
\input{0_abstract.tex}   
\end{abstract}


\maketitle


\input{1_intro.tex}
\input{2_overviewObfus.tex}
\input{3_approach.tex}
\input{4_eval.tex}
\input{5_discussion.tex}
\input{6_relatedwork.tex}
\input{7_concl.tex}

\section*{Availability}
\sh and all data sets that we used as a foundation for our evaluation are freely available at: \url{https://github.com/stg-tud/StringHound}.

\begin{acks}
This work was supported by the Hessian LOEWE initiative within the Software-Factory 4.0 project, by the DFG as part of CRC 1119 CROSSING, by the German Federal Ministry of Education and Research (BMBF) as well as by the Hessen State Ministry for Higher Education, Research and the Arts (HMWK) within their joint support of the National Research Center for Applied Cybersecurity ATHENE.
\end{acks}

\bibliographystyle{ACM-Reference-Format}
\bibliography{references}

\appendix
\input{8_appendix.tex}
\end{document}

%% file: macros.tex
\theoremstyle{definition}

\newcommand{\code}[1]{{\texttt{#1}}}
\newcommand\OPAL[0]{OPAL\xspace}
\newcommand\toolname[0]{StringHound}
\newcommand\sh[0]{\iffinal \toolname \else \toolname\fi\xspace}
\newcommand\LOI[0]{\emph{LoI}\xspace}

\definecolor{lightgray}{gray}{0.85}

\newcommand\greybox[1]{%
\vspace{0.7mm}
\begin{obs}
#1
\end{obs}
\vspace{0.7mm}
}

\makeatletter
\newcommand{\removelatexerror}{\let\@latex@error\@gobble}
\makeatother

\definecolor{gray}{gray}{0.7}
\definecolor{light-gray}{gray}{0.9}
\definecolor{KWColor}{rgb}{0.37,0.08,0.25}
\definecolor{CommentColor}{rgb}{0.133,0.545,0.133}
\definecolor{StringColor}{rgb}{0,0.126,0.941}

%% file: 0_abstract.tex
String obfuscation is an established technique used by proprietary, closed-source applications to protect intellectual property. Furthermore, it is also frequently used to hide spyware or malware in applications. In both cases, the techniques range from bit-manipulation over XOR operations to AES encryption. However, string obfuscation techniques/tools suffer from one shared weakness: They generally have to embed the necessary logic to deobfuscate strings into the app code.

In this paper, we show that most of the string obfuscation techniques found in malicious and benign applications for Android can easily be broken in an automated fashion.
We developed \sh, an open-source tool that uses novel techniques that identify obfuscated strings and reconstruct the originals using slicing. 

We evaluated \sh on both benign and malicious Android apps. In summary, we deobfuscate almost 30 times more obfuscated strings than other string deobfuscation tools. Additionally, we analyzed 100,000 Google Play Store apps and found multiple obfuscated strings that hide vulnerable cryptographic usages, insecure internet accesses, API keys, hard-coded passwords, and exploitation of privileges without the awareness of the developer. Furthermore, our analysis reveals that not only malware uses string obfuscation but also benign apps make extensive use of string obfuscation.

%% file: 1_intro.tex
\section{Introduction}
\label{intro}

Obfuscation protects applications against abusive practices (e.g., repackaging) but also hides malicious intent (e.g., malware)~\cite{dong_2018_understanding, mirzaei2018androdet}. It significantly impedes the analysis of apps 
to check their compliance with privacy regulations or to inspect them 
for detecting malware~\cite{rasthofer_harvesting_2016,menezes_2017_detecting}. In particular, \textit{string obfuscation}, which is applied by most existing obfuscators \cite{dexguard2015guardsquare, allatori2019obfus, dasho2019obfus, stringer2014obfus, zkm2019obfus}, can hide paths, URLs, and intents to track user activities, thus compromising the user's privacy, or opening shells for remote command execution to execute a malicious payload. 

Opposing prior work~\cite{dong_2018_understanding, mirzaei2018androdet, wang2017changed}, which have stated that strings are often not obfuscated in the wild,
in this paper, we provide strong empirical evidence (cf. Section~\ref{prevalence}) that it is widely
used in both malicious and benign apps. The usage of string obfuscation in the benign apps is to a significant extent due to integrated ad libraries -- hence, even the app developer may not be aware of their presence. 
Under these conditions, approaches that analyze plain strings~\cite{zhao2019geo, zuo2019does, nan2018finding, pan2017dark, mariconti2016mamadroid, fratantonio2016triggerscope, wong2016intellidroid} are deemed to be ineffective, and techniques for automatically uncovering obfuscated strings are highly needed.

Given that the deobfuscation logic usually is part of the application~\cite{wermke2018large}, an analyst can try to debug or to run the application with a monkey script. However, such "brute-force" testing has serious drawbacks. First, given that there is no guarantee that all execution paths are covered and that the appropriate execution point to deobfuscate a string is unknown, the latter maybe not triggered. Second, obfuscated applications could detect the debugging mode and avoid to execute the deobfuscation~\cite{rasthofer_harvesting_2016} since deobfuscation functionality is often protected by guards trying to defend against artificial runtime environments~\cite{vidas2014evading}.

Several dedicated approaches~\cite{dexoracle2018fenton,simplify2019fenton,jmd2019contra,dex2jar2019decrypt,rasthofer_harvesting_2016,bello2018ares,wong2018tackling,zhou_2015_harvesting} have been proposed to address obfuscation. But, they suffer from limited scalability and generality.
Many of the existing approaches~\cite{rasthofer_harvesting_2016,bello2018ares,wong2018tackling} typically alter \code{if} statements of the target program and then run the code with all combinations of values to circumvent defenses and force the execution of all branches.
Given that many obfuscators perform automatic string obfuscation on millions of apps, the above approaches are not suited for large-scale analyses.
The approach by Zhou et al.~\cite{zhou_2015_harvesting} slightly reduces the number of executions, but at the cost of generality, as its emulator is fitted to string operations only.
In fact, to the best of our knowledge, all works lack a systematic analysis of existing automatic obfuscators and their scope.

To address the above problems of state-of-the-art deobfuscators, we propose
\sh, a novel string deobfuscation technique for automatically obfuscated strings in Java bytecode. \sh generalizes to different string obfuscations and executes only the code necessary for their deobfuscation to ensure scalability.

Additionally, we performed a comprehensive study of existing obfuscation techniques and used the gained insights to guide the design of \sh to ensure generality. Therefore, we systematically studied how strings are obfuscated in ad libraries (cf. Section \ref{caseStudy}). These libraries often employ string obfuscation
~\cite{stevens2012investigating, razaghpanah2018apps, demetriou2016free, son2016mobile, continella2017obfuscation}
and are, hence, a good source for systematically surveying string obfuscation techniques used in the wild.
To ensure that only code necessary for deobfuscation is executed, 
\sh (a) locates the usage of obfuscated strings within the application code, (b) extracts the deobfuscation logic alongside with all the context it needs to perform, and (c) executes the extracted code directly on a Java Virtual Machine (JVM) to yield the plain-text versions of obfuscated strings.

For a fast location, we propose two classifiers, one that uses decision trees \cite{quinlan1986induction} to identify potentially obfuscated strings, and another one that uses Spearman correlation \cite{myers2004spearman} to identify code of deobfuscation methods.
Given that the deobfuscation logic usually is part of the application~\cite{wermke2018large}, we propose a specifically targeted slicing technique that includes all program statements which affect the state of an obfuscated string sink located within a given method. Additionally, \sh extracts the execution context of the deobfuscation logic and injects the slice into it. Through the injection of the slice, countermeasures, potentially introduced by obfuscators, are rendered ineffective. Finally, \sh executes the resulting slice within the extracted context to obtain deobfuscated strings. 

We evaluated \sh and four state-of-the-art deobfuscation tools~\cite{dexoracle2018fenton,simplify2019fenton,jmd2019contra,dex2jar2019decrypt} by applying them to a set of apps that we obfuscated with 21 different techniques. 
The evaluation shows that \sh{} yields significantly better results than other tools.
We also applied \sh to
four sets of benign and malicious real-world apps:
(a) a random sample of 100,000 apps,
(b) popular apps based on AndroidRank Top~500~\cite{androidrank2019ranking},
(c) malware from Contagio~\cite{contagio2019dump}, and
(d) apps from the Google Play Store in 2018 classified as malicious by VirusTotal.
\sh{}'s classifiers 
were key to enabling a study of more than 100,000 apps by using them to filter out apps that do not contain any obfuscated strings to avoid unnecessary slicing and deobfuscation steps. 
A brute-force approach that tries to deobfuscate each string in all apps would be infeasible, given that an app such as, WhatsApp~\cite{whatsapp2018messenger} contain 17,176 strings.

Our study shows that string obfuscation is used at least 12 times more often than previous studies stated~\cite{dong_2018_understanding, mirzaei2018androdet, wang2017changed}. We categorize our findings and give insights on how string-obfuscation is used in different kinds of apps. 
Besides expected results, e.g., obfuscated URLs and commands in malware sets, we surprisingly found that 76\% of the 100,000 apps contain obfuscated strings. An in-depth analysis revealed that several strings are commonly found in ad libraries integrated into apps. 
Moreover, we identified two apps in the Top 500 set that conceal suspicious behavior through string obfuscation. They collect sensitive information from a user's mobile phone, such as call logs and location information, to build a user profile for tracking. Furthermore, they also check for the \code{SuperUser.apk}, which grants root access to the mobile phone. 
These apps are installed over 20 million times and are not flagged as malicious by VirusTotal~\cite{virustotal2019search}.\\

\noindent
In summary, this work makes the following contributions:

\begin{enumerate}[topsep=0pt,parsep=0pt,itemsep=2pt]
\item A study which identified 21 unique string obfuscation techniques used by state-of-the-art obfuscators (Section~\ref{caseStudy}).

\item Two novel techniques for locating string obfuscation (Section~\ref{features}~\&~\ref{methodClassifier}). 

\item \sh{}\footnote{https://github.com/stg-tud/StringHound}, an novel open-source string deobfuscator that integrates the proposed classifier and slicing techniques. It yields significantly better results than other deobfuscators on our evaluation data set and renders intra-procedural obfuscation techniques ineffective. 

\item A study of string obfuscation in four real-world data sets (Section~\ref{categories}) containing more than 100,000 apps and providing valuable insights into the prevalence of obfuscation usage in the wild.

\end{enumerate}

%% file: 2_overviewObfus.tex
\section{Existing String Obfuscation Techniques}
\label{caseStudy}

We systematically analyzed string obfuscation in ad libraries, since these libraries have been shown to use it in various forms and quantities~\cite{stevens2012investigating, razaghpanah2018apps, demetriou2016free, son2016mobile, continella2017obfuscation} 
The knowledge gained from the following analysis was used as a basis for designing our approach and conducting controlled experiments for evaluation purposes.

\begin{table}
    \centering
    \caption{String Obfuscation Techniques in Ad Libraries}
    \footnotesize
    \begin{tabularx}{\columnwidth}{llllc}
        \toprule
       \textbf{Example Package} & \textbf{Cipher} & \textbf{Encoding} & \textbf{Countermeasure} & \textbf{Count}\\
       \midrule
       com.chamspire & & B64  & & 9\\
       com.intentsoftware & & B85 & & 3\\
       com.ironsource & & custom & & 16\\
       com.youmi & & custom & & 6\\
       com.adcolony & & URL & & 6\\
       a.a.a & AES & & SO & 3\\
       com.google.android &AES \& Bit & B64 & SI & 22\\
       cn.pro.sdk & Bit & & BA & 13\\
       br.com.tempest & Bit & & SW mod. key & 3\\
       com.applovin & Bit & & Key in BA & 25\\
       br.com.tempest & Bit & & Key is SC (KSC) & 3\\
       com.tnkfactory & Bit & & OI & 9\\
       br.com.tempest & Bit & & SC & 3\\
       com.google.android & Bit & & ST & 9\\
       br.com.tempest & Bit & & SW & 3\\
       com.apptracker & Bit & & TK & 9\\
       com.adlib & Bit &  & TM & 13\\
       com.mnt & Bit & B64 & Key is idx of BA & 9\\
       com.waystorm.ads & Bit & B64 & KMC & 31\\
       com.vpon.adon & DESede & & & 28\\
       com.mt.airad & DESede & B64 & & 25\\
       \bottomrule
    \end{tabularx}
    \label{tab:obfuscationTechniques}
\end{table}
\emph{Methodology.}
As there is no publicly available set of ad libraries that use string obfuscation, we sampled our own collection of ad libraries by analyzing apps which integrate them. 
First, we collected a list of package names of frequently used ad libraries~\cite{li2015investigation-tr} and a list of URLs of ad networks \cite{ad2019networks}. 
We reversed the internet domain name~\cite{oracle2018naming} (e.g., youmi.net $\Rightarrow$ \code{net.youmi}) of the URLs to guess package names of ad libraries. 
Next, we searched for code with the respective package names by analyzing 100,000 randomly sampled apps from AndroZoo \cite{hurier2016lack} and found 640 unique ad libraries distributed across 81,008 individual apps.

To identify string obfuscation techniques, we manually inspected obfuscated strings and analyzed their flows until they reached methods that are not modifiable by the obfuscator (e.g., \code{System.println}). Nevertheless, we did not focus only on string constants because, in the obfuscated form, they are often also stored in byte arrays~\cite{schrittwieser2016protecting}. Hence, we considered any data structure which can be used to hide string representations and refer to such data structures in the following as obfuscated strings.
 
During our analysis, we classified a string as not obfuscated, if it flows, without any modification, into an unmodifiable method~\footnote{Most obfuscators produce strings with unreadable symbols and, therefore, contain no words.}. Additionally, if the string contains multiple words found in a dictionary or matches a known format (e.g., XML), it is not classified as obfuscated. 
In the case of other data structures, we considered all bit operations that are performed on constant values to be an indication for string obfuscation.
For each obfuscated string, we then manually analyzed the code that deobfuscates the string to determine the used technique.

\paragraph*{Overview of identified techniques.}
Using our methodology, we identified 21 unique string obfuscation techniques found in 236 ad libraries shown in Table \ref{tab:obfuscationTechniques}. 
Among the identified techniques are also those used by the state-of-the-art obfuscation tool manufacturers such as, DexGuard~5.5.41~\cite{dexguard2015guardsquare}, Allatori~6.8~\cite{allatori2019obfus}, DashO~9.2~\cite{dasho2019obfus}, Stringer~3.0.5~\cite{stringer2014obfus}, ZKM 12.0~\cite{zkm2019obfus}, and Shield4J~\cite{shield4j2020}.
For each technique, we show the cipher, the encoding, and countermeasures used to make detection by static/dynamic analyses more difficult. Additionally, we list the distribution (Count) of each string obfuscation technique across all 236 ad libraries (incl. duplicates).
The used ciphers are bit manipulations such as XOR operations (Bit), DESede, AES, and the combination of bit manipulation and AES. The encodings consist of Base64 (B64), URLEncoder (URL), Base85 (B85), or custom encodings (custom), e.g., using a BigInteger with base 33, or splitting a string and concatenating the characters at the beginning and the end of the new string. 
We identified the following countermeasures: 

\emph{Serialized Object (SO):} One technique loads a serialized object at runtime that implements a deobfuscation method. Subsequently, it must be called through reflection to deobfuscate a string. This technique evades deobfuscators that rely exclusively on identifying and executing deobfuscation methods. 

\emph{Static Initializer (SI):} The static initializer computes the deobfuscation key. This practice evades deobfuscators who extract the logic of only one particular method for execution.

\emph{Byte Arrays (BA):} Two of the analyzed techniques use byte arrays to hide the representation of obfuscated strings and, thus, evade deobfuscators that rely on this representation.

\emph{Switch Statements (SW):} Two techniques use a switch statement in a loop to deobfuscate a different string in each loop iteration. Both store the resulting strings in an array, and each method accesses this array. These techniques evade deobfuscators that search for an explicit deobfuscation method.

\emph{Stack Calls (SC):} Two techniques hard-code the calling context (e.g., method name and class name) of the deobfuscation method.
While one technique checks the calling context in a conditional statement, the second one uses the calling-context information as part of the deobfuscation key. Both techniques evade deobfuscators that execute the deobfuscation logic without a specific context.
However, only the second one enforces the extraction of the context for slicing approaches because it is a direct part of the deobfuscation. 

 \emph{Object Initializer (OI):} One technique deobfuscates strings by inserting a specific class whose constructor initializes the deobfuscation key. Subsequently, a method of the constructed object deobfuscates all strings which were obfuscated with the initialized key. 
This technique evades deobfuscators that execute only static methods.

\emph{Stream Transfer (ST):} Hidden channels are used to transfer obfuscated strings to deobfuscation methods. For instance, one obfuscator transfers the obfuscated string via input/output streams to its deobfuscation method. This technique evades deobfuscators that track obfuscated strings and would, therefore, miss data flows arising from streams.

\emph{Two Keys (TK):} Two different keys are used for string deobfuscation. 
This usage evades deobfuscators that try brute force guessing of one key to uncover obfuscated strings.

\emph{Two Methods (TM):} Two methods are used for string deobfuscation.  
This usage evades deobfuscators that execute only one deobfuscation method to uncover obfuscated strings.

\emph{Key Management Calls (KMC):}  One technique initializes deobfuscation keys directly before their usage by using object fields. This technique hinders deobfuscators that do not handle the initialization of fields.

As depicted in Table \ref{tab:obfuscationTechniques}, different combinations of ciphers, encodings, and countermeasures are used as techniques for string obfuscation. 
We refer to these combinations as obfuscation schemes.
Some of the techniques are used in state-of-the-art commercial obfuscation tools, and developers most commonly use these tools to obfuscate strings in Android and Java apps.
The findings of this study are surprising as none of the identified techniques requires a broader focus than the one described above. 

\greybox{
All analyzed obfuscation schemes are initialized within the class containing the deobfuscation methods. Thus, no heavyweight inter-procedural analysis seems necessary for our data set.
}

Next, the gained knowledge of these identified schemes is used to evaluate our approach. 
Therefore, we obfuscated samples with all schemes by using either an acquired tool or a re-implemented version of the scheme. 
The re-implementation was achieved by manually reversing the deobfuscation logic found in obfuscated apps. 
For instance, if the deobfuscation logic used a Base64 decoding followed by an AES decryption, we first encrypted all strings with AES and then encoded the result using Base64.

%% file: 3_approach.tex
\section{The StringHound Approach}
\begin{figure}[tb]
    \centering
    \includegraphics[width=\columnwidth]{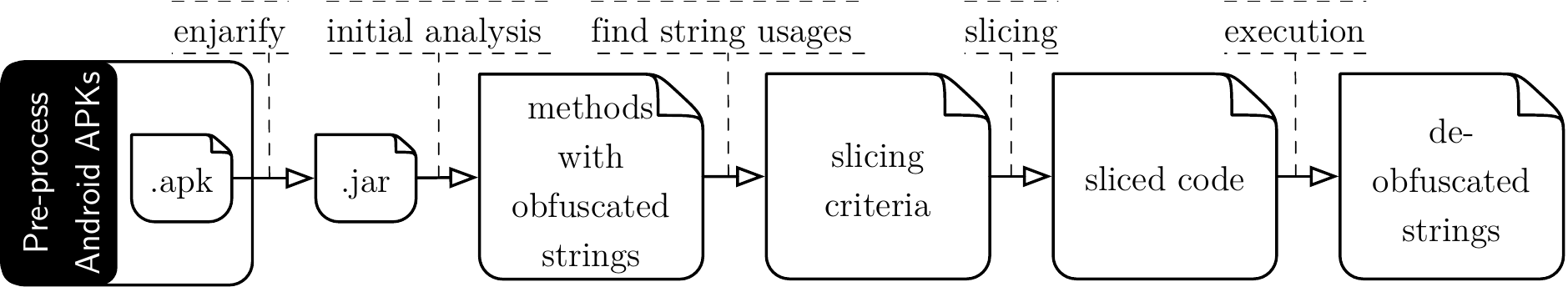}
    \caption{Overview of \sh's Approach}
    \label{fig_process}
\end{figure}

\sh processes Java bytecode in five steps. 
\figurename~\ref{fig_process} shows a high-level view of this process.
First, when we analyze an Android Package (APK), we transform its Dalvik bytecode to Java bytecode and process the result with our analysis.
Second, to reveal obfuscated strings, we need to identify the methods that potentially use them.
For locating usages of obfuscated strings, we propose two complementary
techniques: a classifier for identifying potentially obfuscated strings (String Classifier), and a classifier for identifying deobfuscation methods (Method Classifier). 
The string classifier operates on characteristics of obfuscated strings by using decision trees. The method classifier matches distributions of instructions from known deobfuscation methods with the Spearman correlation.
Third, we find the starting point for the slicing (slicing criterion) in the methods that contain the usage of the obfuscated strings. 
Forth, we use a specifically targeted slicing technique that computes all program statements that affect the state of a given slicing criterion.
Finally, \sh injects the slice into the execution context of the deobfuscation logic. Afterward, it executes the resulting slice to obtain deobfuscated strings. The injection of the slice into the context renders countermeasures introduced by obfuscators ineffective. Our detailed description of \sh shows the design decisions taken to address the obfuscation schemes presented in Section~\ref{caseStudy}.

\label{approach}
\subsection{Classifiers}
In this section, we present our classifiers and evaluate their precision and recall. In Section \ref{prevalence}, we provide empirical evidence that they are both needed.

\subsubsection{Training \& Evaluation Data Set}
\label{fdroidData}
For the training and evaluation of our classifiers we downloaded the newest versions of all 1,879 apps from F-Droid~\cite{fdroid2019store}. 
We chose F-Droid because it only consists of open-source software. 
Hence, string obfuscation does not make sense and is also not applied. 
This data set is used as ground truth of plain strings and methods
which do not contain any deobfuscation logic. These two properties enable the obfuscation without dealing with the influence of previously existing obfuscation artifacts. 
The standard configuration of the Android build process might use name obfuscation due to the integration of ProGuard. However, ProGuard does not use string obfuscation~\cite{provsdex2020}.

\subsubsection{String Classifier}
\label{features}

To train the classifier, we generated a data set of 1,918,687 obfuscated from the same amount of non-obfuscated strings. The set of non-obfuscated strings was extracted from the F-Droid data set (cf. Section \ref{fdroidData}). We applied all 21 obfuscation schemes identified in our case study (Section~\ref{caseStudy}) to non-obfuscated strings of F-Droid applications to obtain obfuscated strings. This effort yields 32,379 obfuscated apps\footnote[1]{We were not able to obfuscate every app with every obfuscator due to version incompatibilities between obfuscators and APKs to be obfuscated.} from which we extracted all strings.
As a result, we acquired significantly more obfuscated strings than plain strings. To avoid a bias towards obfuscated strings, we took all strings from the plain apps and randomly selected the same number of strings from 
the obfuscated ones.

\paragraph*{Approach}
We extracted 49 different features from the collected strings and trained a model using Weka's REPTree. This enables a fast comparison of the features by building a regression/decision tree using the most discriminatory features to check the most relevant ones first. However, REPTree considers all these features before it comes to a final decision for a given string. The authors checked that the classifier considers all features for its final decision by manually traversing the decision tree.

\begin{table}[b]
\centering
    \caption{Feature List for the Detection of Obfuscated Strings}
    \footnotesize
\begin{tabularx}{\columnwidth}{ll}
\hline
\textbf{Category}  & \textbf{Description}\\ \hline
Format    & - User Agents\\
& - URLs\\
& - Character set of regular expressions\\
& - Network protocols (e.g., WiFi)\\
& - Common OS commands\\
& - JSON format\\
& - Encodings (e.g., UTF-8)\\
& - E-Mail address\\
& - DTD\\
& - HTML Colors\\
& - Class Path\\
& - SQL Queries\\
& - Keywords for seven programming languages\\
& - Country names\\
& - XML\\
& - IP\\
& - HTTP state\\
& - Date\\
& - Numeric\\
& - Cryptographic primitives\\
& - Mobile phone brands\\
& - HTML special characters (e.g., uuml)\\
& - String-encoded certificate\\
& - String-encoded Android certificate\\
& - Private/Public key\\
& - String signatures of social network apps\\
& - String-encoded images (e.g. JPEG)\\ \hline
Statistical Tests & - Tests if all chars in the given string are equally,\\
& distributed indicating a random distribution. \\
& - The average distribution which is close to the\\
& Gaussian distribution for plain strings,  \\ 
& - The normalized entropy of the strings,\\ \hline
AndroDet\cite{mirzaei2018androdet} & - Number of equals\\
& - Number of dashes\\
& - Number of slashes\\
& - Number of pluses\\
& - Sum of repetitive characters\\ \hline
Compression Rate & - The rate of the GZIP compression\\ \hline
Cryptography Library  & - The string is used in a known crypto library\\ \hline
Dictionary Words & - The shortest word length\\
&  - The largest word length\\
& - The number of words\\
& - The number of unique words from a\\
& multiple language dictionary\\ \hline
String Characteristics & - Number of vocals\\
& - Number of consonants\\
& - Number of digits\\
& - Number of characters\\
& - Number of unique characters\\
& - Number of non letters\\
& - Maximum number of consecutive characters\\
& - Maximum occurrences of the same character
\end{tabularx}
 \label{tab:features}
\end{table}

In the following, we describe the used features from Table~\ref{tab:features} grouped by their category.

\emph{Format:}
In the study presented in Section \ref{caseStudy}, we observed that obfuscated strings often contain non alphanumeric characters. Nevertheless, we cannot classify a string as obfuscated just because it contains non alphanumeric characters -- plain strings of certain formats may also use these characters. To avoid matching such plain strings, we use various patterns to discern format usages such as XML (e.g. \code{</th>}) and HTML colors (e.g. \code{\#FFAE40}) in the feature vector. These flags are used to give the model a hint that the analyzed string might not be obfuscated. However, these hints should not be confused with filtering, as they are only a part of the classifier's decision.

\emph{Statistical Tests:}
Previous statistical analyses of encryption mechanisms~\cite{kahn1996codebreakers,cryptanalysis2019lyons,lyda2007using} show that obfuscated strings often have a random (close to equal) distribution of characters. We use random distribution as a discriminating feature to distinguish between obfuscated and other strings with special characters. 
To check whether the distribution of the characters in a string is random, we use three different measures because each one is suited for different scenarios we encountered.
With the \emph{Chi-squared} test, we measure the deviation of the characters from the equal distribution of these characters, since randomized characters are often equally distributed. With the deviation from the \emph{average distribution}, we measure whether the given string belongs to a language or whether the characters were only rotated (e.g. caesar cipher~\cite{kahn1996codebreakers}). \emph{Normalized entropy} was previously~\cite{lyda2007using} used to identify encrypted malware. We reuse it to identify encrypted strings.

\emph{AndroDet:}
We use the number of equals, number of dashes, number of slashes, number of pluses, and the sum of repetitive characters from the feature list of  AndroDet~\cite{mirzaei2018androdet} which are used to identify if an app uses string obfuscation. However, AndroDet averages these features over all strings in an app and is therefore not able to classify individual strings. 

\emph{Compression rate:}
Obfuscated strings may be confused with compressed data, such as images compressed using JPEG and stored in strings. To identify those strings, we compressed them and compared the resulting length against the original length; the resulting length changes if the original content is not already compressed~\cite{cryptanalysis2019lyons}. 

\emph{Cryptographic libraries:}
Cryptographic libraries use byte-encoded strings to initialize their algorithms, and this may cause false positives because they are similar to obfuscated strings. To avoid matching such encoded strings, we check if a string usage is contained in a known cryptographic library.

\emph{Dictionary words:}
The study in Section \ref{caseStudy} revealed that obfuscated strings contain only a few words or none at all.
We use a dictionary to check whether a string contains words or identifiers~\cite{enslen2009mining}. 
To match words from foreign languages (e.g., Chinese) that do not use separators such as white spaces, we apply Lucene's \textit{ICUTokenizer} word splitting approaches. 
Furthermore, to match strings consisting of concatenated words (e.g. \textit{getLength}), we use Samurai~\cite{enslen2009mining}, which splits identifiers by camel case and frequently used words.

\emph{String characteristics:}
Finally, we extract eight features related to character distributions, e.g., character counts, digits.\\

\textbf{\textit{Evaluation} }
We use 80\% of our data set for training and testing and 20\% for validation. To train and test the model, we use a 10-fold cross-validation measure.
Our validation data revealed a precision of 98.79\% and a recall of 89.75\%. 
We identified two root causes for false negatives. The first cause is that obfuscated strings accidentally contain valid words (this is exacerbated by languages where a single character can be a valid word, e.g. Chinese). 
The second, more prevalent cause is obfuscated strings that consist of digits since these frequently occur in plain text strings as well. 

\subsubsection{Classifier for Deobfuscation Methods}
\label{methodClassifier}

The string classifier may miss obfuscated strings that are hidden in other data types, e.g., strings represented as byte arrays (cf. Table \ref{tab:obfuscationTechniques} BA).
To address this problem, we train a second classifier that identifies deobfuscation methods. We use the identified methods from Section~\ref{caseStudy}.

\paragraph*{Approach}

We postulate that deobfuscation methods use certain instructions more often than ordinary methods. This idea is inspired by statistical analysis of English text, which, e.g., contains a high number of the character 'e' \cite{cryptanalysis2019lyons}. Likewise, deobfuscation methods may use the XOR instructions more frequently than ordinary methods. 
To this end, we extract all instructions used in deobfuscation methods of the identified schemes (cf. Section~\ref{caseStudy}). 
The extraction of the instructions is performed using the \emph{Structure-preserving Representation (SPR)} \cite{glanz2017codematch}. 
This representation preserves the structural tokens of a method's instructions but abstracts away information that gets changed in obfuscated code and, thus, would produce noise for the classification, e.g., all name and type information that does not occur in the Android standard library is removed.
We compare the SPR-token distribution of our set of deobfuscation methods with the ones found in apps using Spearman's correlation to identify similar methods. 
This comparison enables the method classifier to handle obfuscation schemes that do not use string representations (e.g. BAs) and identify not only exact matches of the token distribution but also variations of it. 
Furthermore, we limit our token extraction to those tokens occurring in known deobfuscation methods; as a result, our method classifier is also able to identify in-lined deobfuscation logic.\\

\textbf{\textit{Evaluation}} The primary purpose of the method classifier is to locate deobfuscation schemes that represent obfuscated strings in other data structures. As reported in Section~\ref{caseStudy}, only two such schemes exist (cf. BA in Table~\ref{tab:obfuscationTechniques}), and these also generate variations of the deobfuscation logic. Nevertheless, to assess the precision and recall of the method classifier, we use not only the schemes which generate variations of known deobfuscation methods as a ground truth but the methods of all the obfuscation tools acquired in Section \ref{caseStudy}. We use methods generated by all tools since the method classifier discriminates all kinds of deobfuscation methods, not only those that handle other data structures than strings.

The two mentioned tools vary the logic of the deobfuscation methods in different ways. 
First, they use random numbers as obfuscation keys. Second, they permute the order of formal parameters or change the method's signature.
Third, they alter the position of code blocks, whose execution order does not matter. Finally, deobfuscation methods may also depend on the context of string usages. 
For instance, if a string is used only once in a class, one tool in-lines the deobfuscation logic at the string usage site; in other cases, this logic is extracted into a separate called method.

To measure the precision of the classifier and recall for each variation, we applied the both obfuscators to the F-Droid data set (cf. Section \ref{fdroidData}).
We were able to generate 2,127 obfuscated apps, at least 1,000 apps for each obfuscator~\footnote{We were not able to obfuscate every app with every obfuscator due to version incompatibilities between obfuscators and APKs to be obfuscated.}. 
The deobfuscation methods in the resulting obfuscated apps constitute our ground truth for measuring recall and precision. 

To extract them, we use information from the mapping files produced by the obfuscator tools for each app. 
Mapping files enable app developers to find the original names in the source code for crash reports using obfuscated names.
Consequently, methods and fields with no entry in the mapping file must have been added by the obfuscator. We add all new methods and also methods that access newly added fields to the ground-truth list. The newly added fields are used to identify in-lined deobfuscation logic, which resides in a previously existing method.

Altogether, we obtain a list of 144,190 methods that contain deobfuscation logic, either in a separate method or in-lined into previously existing methods.
The comparison of this list with the method classifier's output shows that it identifies the variants of deobfuscation methods generated by the two subject obfuscator tools with a precision of 
99.66\% and a recall of 97.42\%.
We conclude that our classifier is very accurate, missing only a few deobfuscation methods. 
A detailed analysis revealed that these methods have in-lined obfuscation logic, 
but already used byte arrays before the obfuscation. 
These previously existing byte arrays add noise to the measured token distribution and weaken the correlation between the method under analysis and our set of known deobfuscation methods.

\subsection{Slicing Relevant String Usages}
\label{sec_idStrings}

A slicing criterion ($s_{crit}$) is any instruction within certain methods, which we call candidate methods, that produces a string value.
A method $m$ is in the set of candidate methods if (a) it contains instructions that consume a char sequence as a parameter (method calls, but also field writes, array stores, and return instructions), called \emph{Locations of Interest ($LoIs$)}, and (b) satisfies one of following conditions:
(i) the string classifier found an obfuscated string in \code{m},
(ii) \code{m} calls a method \code{n}, which the method classifier identified as a deobfuscation method,
or (iii) \code{m} is itself classified as a deobfuscation method (in-lined deobfuscation logic). 

Since the classifiers identify neither \LOI{}s nor slicing criteria, we have to search for them in the candidate methods. We use \OPAL~\cite{Eichberg_Hermann_2014} to find all instructions that operate on values of type \code{CharSequence}, or a subtype thereof, in particular \code{java.lang.}\code{String}.
All $s_{crit}$ are expressions that result in strings which are afterward passed to some \LOI.
Given a candidate method that contains \LOI{}s, we identify all $s_{crit}$ while ignoring constant string expressions.

Our slicing algorithm performs backward slicing with forward-phases to collect all instructions necessary for the execution of other relevant instructions. For instance, if the slice contains a \emph{new} instruction, we also collect the corresponding constructor invocation. Additionally, if several potential sources for a given string parameter are present, we start form each of them as separate slicing criterion. 

For example, Listing~\ref{lst:label} shows two sources of \textit{msg}~(Line~2) corresponding to the two branches of the tertiary operator~(Line~1), which load either \code{"US()"} or \code{"INT()"}. In such cases, \sh would start the slicing process for each source.
\begin{lstlisting}[aboveskip=0.40 \baselineskip,belowskip=0.40 \baselineskip,caption={Example with Two Sources},label={lst:label}]
String msg = simCountryIso().equals("US") ? US() : INT();
invoke("+01234", msg); 
\end{lstlisting}

\subsection{Our targeted Slicing}
Our slicing technique (cf. Algorithm~\ref{alg_slicing}) is inspired by traditional slicing algorithms (cf. Binkley et al.~\cite{binkley1996program}), and implemented using OPAL~\cite{Eichberg_Hermann_2014} with definitions (cf. Aho et al.~\cite{aho1986compilers}) of the functions defined in Table~\ref{tab:definitions}.

\begin{table}[b]
    \centering
    \caption{Definitions of Helper Functions for the Algorithm}
    \footnotesize
    \begin{tabularx}{\columnwidth}{lll}
        \toprule
    def & $Instr \rightarrow \mathcal{P}(Var)$ & variables defined by an instruction\\
    use & $Instr \rightarrow \mathcal{P}(Var)$ & variables used by an instruction\\
    du & $Var \times Instr \rightarrow \mathcal{P}(Instr)$ & definition-use instructions\\
    ud & $Var \times Instr \rightarrow \mathcal{P}(Instr)$ & use-definition instructions\\
    cd & $Instr \rightarrow \mathcal{P}(Instr)$ & transitive control dependency instructions\\
    br & $Instr \rightarrow \mathcal{P}(Instr)$ & set of backwards reachable instructions\\
       \bottomrule
    \end{tabularx}
    \label{tab:definitions}
\end{table}

\label{sec_slicing}
    \begin{algorithm}[b]
    \footnotesize
        \KwIn{\hspace{0.8em}$m$ a method with a body \\
        \hspace{4em}$I$ the instructions of the method $m$ \\
        \hspace{4em}$g$ the CFG of $m$ where each $i \in I$  \\
        \hspace{5em}corresponds to one node $n \in N$ of $g$ \\
        \hspace{4em}$LoI \in I$ the location of interest \\
        \hspace{4em}$s_{crit} \in I$ the slicing criterion}
        \KwOut{$N_{slice} \subseteq I$ }
        
        $N_{slice} := \{\}$\\
        $W := \{s_{crit}\}$ \\
        $cd_{crit} := cd(s_{crit}$) \\  
        $br_{LoI} := br(LoI)$ \\
        \While{$W \neq \emptyset$}{
            $currInstr$ $:= head(W)$ \\
            $W := W~\backslash~currInstr $ \\
            \If{currInstr $\notin N_{slice}$} {
                $N_{slice}$ := $N_{slice}$ $\cup$ \{ $currInstr$ \}\\ 
                $D := \{d~|~x \in use(currInstr)\ \wedge d \in ud(x,currInstr)\}$ \\
                $cd_{currInstr} := cd(currInstr)~\backslash~$cd$_{crit}$\\
                $U := \{u~|~ x \in def(currInstr) \wedge u \in du(x,currInstr)$ \\
                $\;\;\;\;\;\;\;\;\; \wedge\ u \in br_{LoI}$\} \\    
                $W := W \ \cup\ $D$\ \cup\ $ $cd_{currInstr}  \cup\ $U \\ 
            }
        }
        \caption{Slicing Algorithm}
    \label{alg_slicing}
    \end{algorithm}

Given a method along with its control-flow graph (CFG), a \LOI{} and a slicing criterion $s_{crit}$, we initialize the worklist $W$ (Line 2 of \figurename~\ref{alg_slicing}) with $s_{crit}$. 
For each instruction in $W$ (Line~6) that is not already part of the slice (Line~8), we perform the following steps:
\begin{enumerate}[wide=0.1cm,topsep=0pt,parsep=1pt,partopsep=0pt,itemsep=2pt]
\item We add the current instruction $currInstr$ to the slice (Line~9). 
\item In the \emph{backward phase} (Line 10), we determine the set $D$ of all definition sites related to $currInstr$, i.e., $D$ consists of instructions that initialize variables used by $currInstr$. 

\item Also in the \emph{backward phase}, we determine the set $cd_{currInstr}$ of instructions on which the current instruction is control dependent on (Line~11). 
From this set, we remove the instructions $cd_{crit}$ that could prevent the execution of $s_{crit}$. 
This \textit{backward phase} adds instructions to $W$ that (in)directly affect $s_{crit}$. 
With this addition, we include condition instructions that do not control the execution of the criterion itself. 
This step is required to, e.g., ensure that loops manipulating byte arrays are added to the slice. 
If the \textit{backward phase} adds instructions that define a new reference-typed variable, i.e., an object, we perform an additional \textit{forward phase} to include those instructions in $W$ that potentially affect the state of the object after its initialization and which are relevant w.r.t.\ the \LOI. 
Hence, we only add instructions that are still backward reachable from the \LOI.

\item In the \emph{forward phase} (Line 12, 13), we determine the set of all instructions $U$ that use a variable defined by $currInstr$ and which are backward reachable from the \LOI{}.
This phase includes all instructions that potentially mutate the state of the defined variable, e.g., filling an array with actual values or calling a method of the object. 

\item In the last step (Line 14), we update $W$ with the three different sets of instructions.
(i) $cd_{currInstr}$  the instructions on which $currInstr$ is control dependent,(ii) $U$ those that use the variable defined in $currInstr$, and (iii) $D$ those that initialize the variables used by  $currInstr$.
\end{enumerate}

\begin{figure}[tb]
\centering
\includegraphics[scale=0.5]{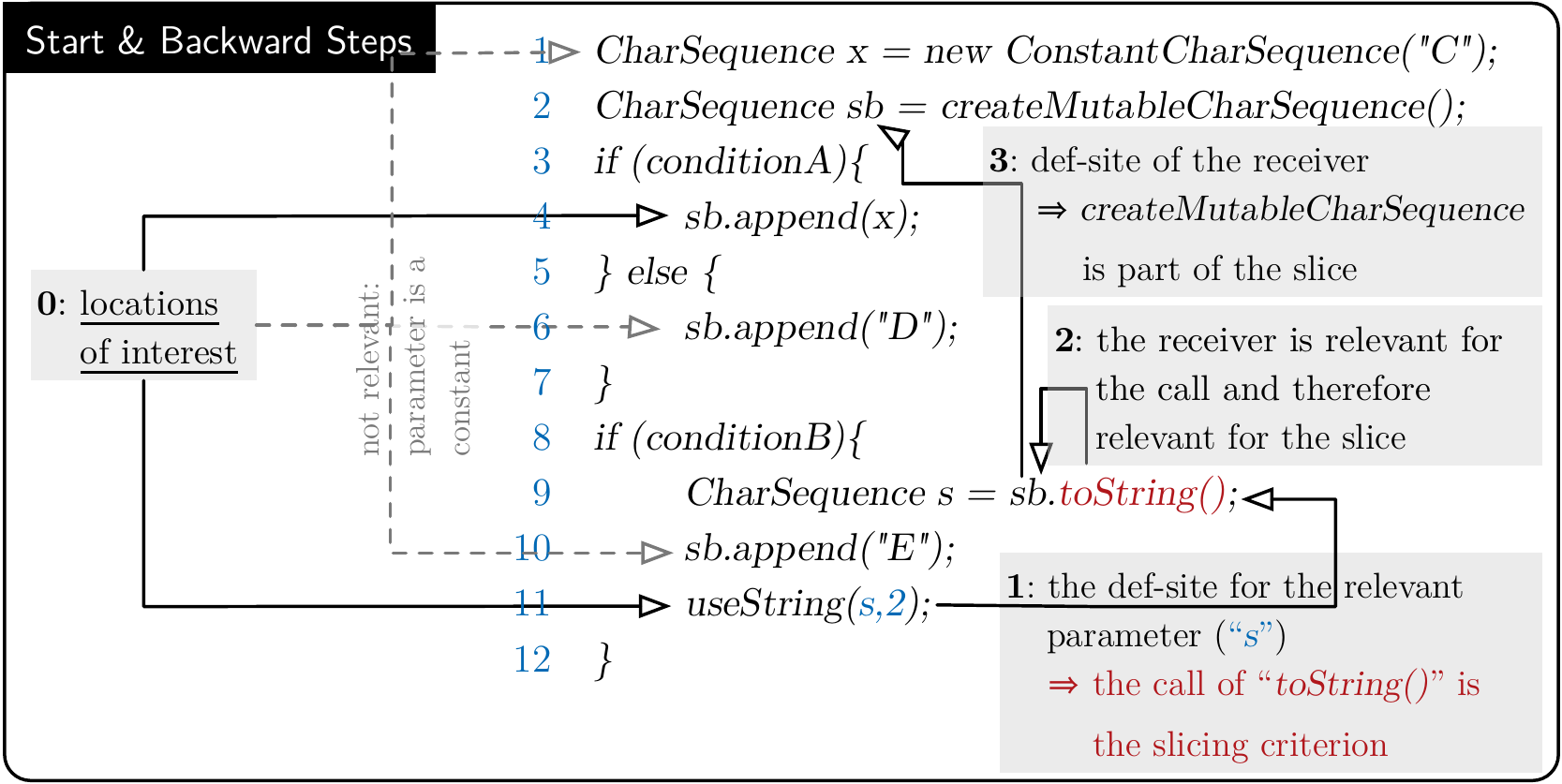}
\caption{Example of the Slicing Process for the Parameter \code{s} of \code{useString}---Showing \LOI{}s and First Backward Phase}
\label{fig_start_and_backward}
\end{figure}

\begin{figure}[tb]
\centering
\includegraphics[scale=0.5]{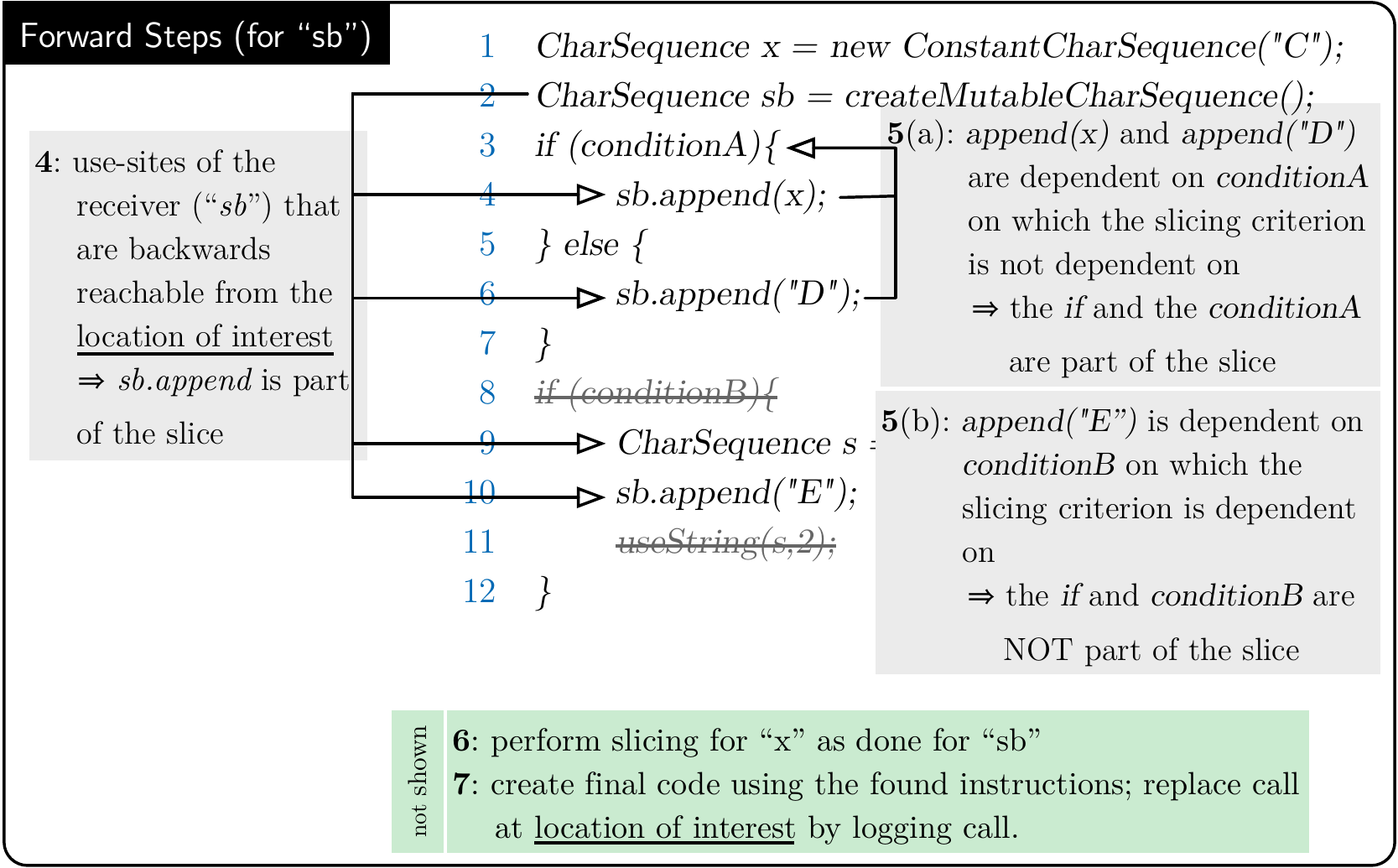}
\caption{Example of the Slicing Process for the Parameter \code{s} of \code{useString} --- Showing the Necessary Forward Phase}
\label{fig_forward_slicing}
\end{figure}

The example in \figurename~\ref{fig_start_and_backward} illustrates the process of determining $s_{crit}$ and the backward phase of the algorithm, divided into 4 steps.
In step~0 we determine the statements that are candidates for \LOI{}s. Here, the constructor (Line 1), the \code{append()} calls (Lines 4, 6, and 10), and the call to \code{useString()} (Line 11) are selected. 
Given a \LOI, we consider the definition sites (def-sites) of the instructions that load the \LOI's string parameters as slicing criteria (without including the \LOI itself). 
For illustration purposes, assume the \LOI being processed is the \code{useString} call (Line~11), the only string parameter \code{s} (the \code{int} parameter is ignored) is defined by the result of the call \code{sb.toString()} (Line 9); hence, this call is our slicing criterion. 
On the contrary, instructions that load string constants are not considered as slicing criteria, e.g., Line 6 is a \LOI, but the instruction that loads the string constant \code{"D"} is not a slicing criterion. The rationale is that in such cases, we are sure that no deobfuscation happens before reaching the \LOI. In \figurename~\ref{fig_start_and_backward} such \LOI{}s are pointed at by dashed arrows.
Therefore, we establish in step~1 that the call \code{useString(s,2)} (Line~11) is the \LOI and \code{toString()} (Line~9) is the $s_{crit}$. 
The first backward slicing phase determines the definition site of the object on which \code{toString()} is called, i.e., \code{sb}, this is step 2. 
Hence, \code{createMutableCharSequence} (Line 2) is added to the slice in step 3. 
The if-condition (Line 8) is not added to the slice because it would potentially prevent the execution of the slicing criterion ($s_{crit}$).

The example continues in \figurename~\ref{fig_forward_slicing}, showing the forward phase of the algorithm.
Here, we perform a forward phase concerning \code{createMutableCharSequence} (Line 2) which does not use variables. 
Thus, no backward step is necessary. 
In step~4 (cf.~\figurename~\ref{fig_forward_slicing}) we identify all use-sites of \code{sb} and, since all of them (Lines 4, 6, 9, and 10) are backward reachable from the \LOI, we add them to the slice. 
In this case, Line 10 is not needed in the slice, because the string returned at Line 9 cannot be mutated afterwards. 
However, our algorithm does not have such knowledge and, therefore, conservatively adds it.
Next, in step~5(a) the algorithm processes the \code{append} calls in Lines 4 and 6 as follows. 
Given that no local variable is defined, there will not be a forward step; however, in the first \code{append} call~(Line 4), we use \code{x} and, therefore, add the defining instruction (Line 1) to the slice. Additionally, the \code{if}-instruction in Line 3 is added to the slice, because both \code{append}s are control dependent on it, but not $s_{crit}$. When we process the \code{append} call in Line 10 in step 5(b) we see that the \code{if}-condition~(Line 8) would possibly prevent the execution of $s_{crit}$ and thus do not add it.

Step 6 is not explicitly shown, as it basically repeats steps 2-5 but starts with the variable \code{x} in Line 4.

To recap, the resulting slice is the entire code from \figurename~\ref{fig_forward_slicing}, except for Lines 8 and 11 in \figurename~\ref{fig_forward_slicing} (which are crossed out). In contrast to our approach, traditional slicing algorithms would include the condition in Line 8 into the slice which could prevent the execution of the relevant code.

In step 7, we create the executable code, including a method to retrieve the value that would have been used at the \LOI{}. 
The following section explains this in more detail.

\subsection{Executing Sliced String Usages}
\label{sec_creationOfMethod}

To obtain the deobfuscated string that in the original application would flow into the \LOI, we extend the slice by a call to a method that logs the string. 
This call effectively replaces the original \LOI with the call to the logging method, which allows us to retrieve the deobfuscated string value. We add a return statement to the slice to ensure that the signature of the sliced method can remain as before. If we need to return a value, we either return \code{null} or the numeric value \code{0}---depending on the declared return type.
Next, we replace the body of the original candidate method with the extended slice; this ensures that the execution context w.r.t.\ the name of the declaring class as well as the name and signature of the method is identical to the original code to evade the countermeasures \emph{SC} and \emph{KSC}.

To execute the sliced method in its context, we have to make the class concrete, if it is abstract. 
Therefore, all abstract methods are made concrete by returning default values of the declared return type.
We rewrite the class so that it extends a superclass that we generate, including the corresponding static initializer and super calls. With this step, we increase the likelihood that the initialization of our class containing the sliced method does not abort with an exception to evade the countermeasures \emph{SI} and \emph{OI}. Recall that we have no means to determine appropriate parameter values that we could use and, therefore, always have to use default values. The generated superclass also implements all methods transitively called by the sliced method. As previously, we return default values if required.

We set up the classpath to include all classes of the original application, except the modified one. 
Additionally, we add the new class as well as our new superclass to ensure that our slice can find any application class used in its code.
As a replacement of the original \code{android.jar}, we use an artificial jar with methods stubs. 
Methods that have to return a value return the type's default value (e.g., \code{null} or 0).
All these transformations together, in combination with our slicing approach, enable \sh to circumvent all obfuscation schemes discussed in Table~\ref{tab:obfuscationTechniques}. 
Even if reflection is used, the slice can be run successfully as long as the targets are part of the execution environment.
Native methods are not included, since most Android applications do not compile their native parts for the x86 architecture on which we perform the slicing. 

Finally, we call the resulting method reflectively using default values for the parameters when necessary. The method will then call our logging method to record the deobfuscated string.\footnote{We can specify a time limit for the slice execution, to cancel long-running slices.}
If the execution of the sliced method crashes, no other slices are affected and only the result of the crashed slice will be missing.
We chose to call the sliced methods with default values because the choice of them is simple and caused no overhead. Nevertheless, our approach does not depend on this choice and can be extended to support more advanced methods for determining the parameter values such as fuzzing.

%% file: 4_eval.tex
\section{Evaluation}
\label{evaluation}

We performed two studies (a) comparing \sh against other string deobfuscators, and
(b) assessing the performance of \sh on real-world apps.

The setup consists of a Server with two AMD(R) EPYC(R) 7542 @ 2.90\, GHz (32 cores / 64 threads each) CPU, and 512~GB RAM.
The analyses were run using OpenJDK 1.8\_212 64-bit VM with 20\, GB of heap memory, and a 5s timeout for a single string deobfuscation.

\subsection{Comparison with Other Deobfuscators}

We evaluated \sh against Dex-Oracle 1.0.5~\cite{dexoracle2018fenton}, 
Simplify 1.2.1~\cite{simplify2019fenton}, 
JMD 1.61~\cite{jmd2019contra}, and DEX2JAR 2.0~\cite{dex2jar2019decrypt}.
To the best of our knowledge, these are the only freely available deobfuscators.

\begin{figure}
    \includegraphics[width=\columnwidth]{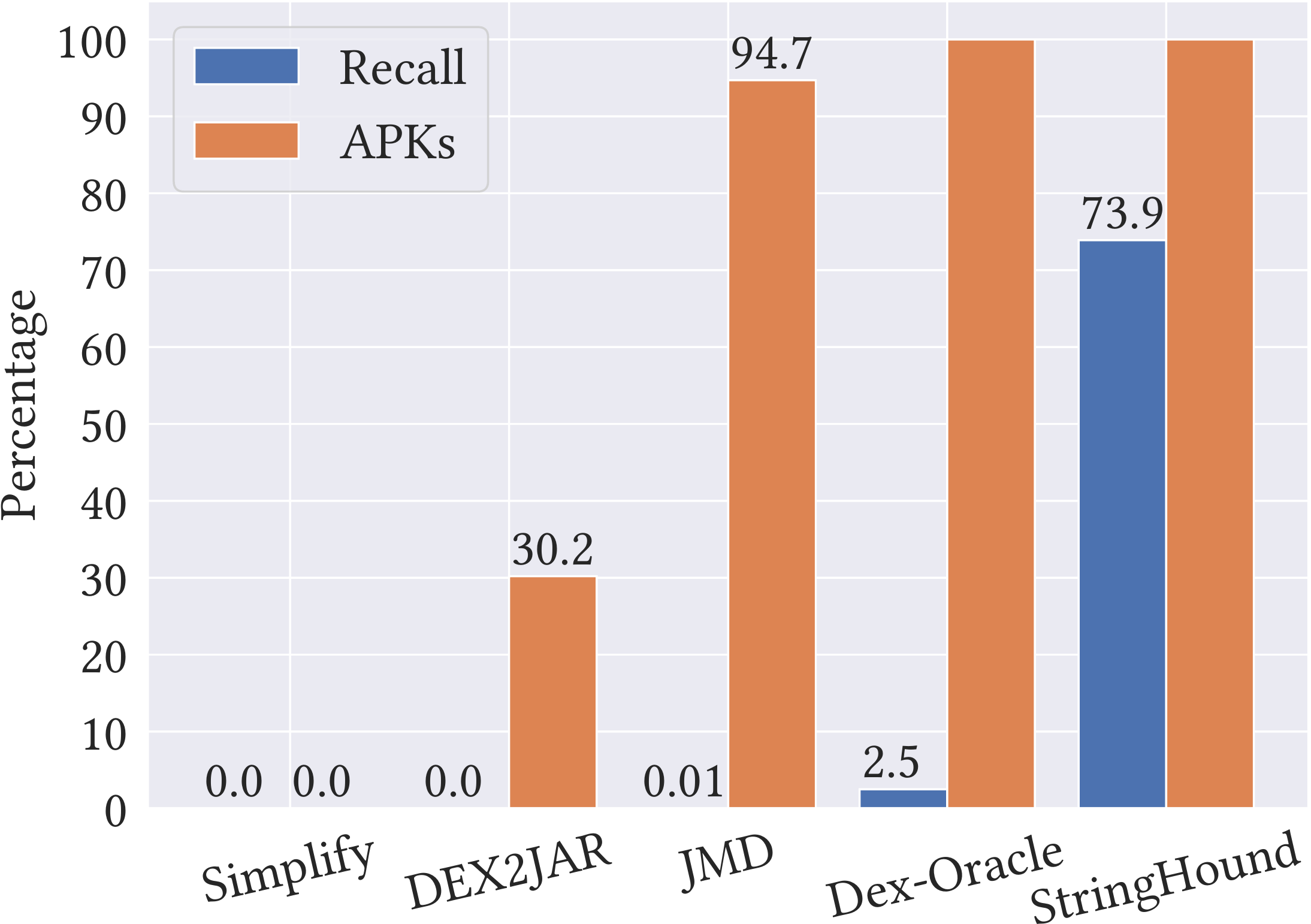}
    \caption{Recall and Successfully Processed APKs}
    \label{fig_accuracy}
\end{figure}

As input for the deobfuscators, we randomly picked 1,000 apps from the data set described in Section~\ref{features} which have not been previously used to train our classifiers.
Two comparison metrics are used: (a) percentage of APKs processed without termination errors; and (b) recall, which we define as the percentage of \emph{unique deobfuscated strings} over  \emph{all unique strings} in the original apps.
The precision metric is discarded since our data set contained only obfuscated strings. Therefore, there cannot be false positives (i.e., plain strings identified as obfuscated). However, \sh's false positive rate is restricted by the false positives produced by the string classifier and the method classifier.
The results are summarized in \figurename~\ref{fig_accuracy}. 
In the following, we discuss each deobfuscator individually.

\emph{Simplify \cite{simplify2019fenton}} applies semantic-preserving transformations to re-engineer the APK's code, such as constant propagation and dead code removal. 
To enable transformations, it executes each method on a custom Dalvik virtual machine and returns a graph with all possible register and class values for every execution path. 
\emph{Simplify} can be used as a deobfuscator in limited cases~\cite{simplify2019fenton}, namely for deobfuscation methods that do not depend on any state and use only constants. In such cases, constant propagation can uncover hidden information. 
Additionally, \emph{Simplify} optimizes all statements also the ones which are not relevant to deobfuscate a string. 
Unfortunately, \emph{Simplify}'s re-engineered APKs were completely broken and could not be analyzed to produce results; hence, \figurename~\ref{fig_accuracy} reports 0\% for both values.

\emph{DEX2JAR \cite{dex2jar2019decrypt}} transforms Dalvik bytecode to Java bytecode. It has a sub-module that executes methods with a certain signature for deobfuscation purposes. Similar to our approach, it executes the code in the JVM. Unlike \sh, DEX2JAR needs the user to provide the deobfuscation method to be executed. We applied our deobfuscation method classifier to each app and used its output as input for DEX2JAR. Providing the same deobfuscation methods as input to DEX2JAR and \sh enables a fair comparison of the two.  
However, DEX2JAR processed only 30\% of the APKs without errors, and it was unable to deobfuscate a single string, resulting in a 0\% recall. This weak results in our empirical study are caused by DEX2JAR's assumption that deobfuscation methods are in the same class as the obfuscated string. 
Moreover, DEX2JAR assumes that all constant values needed for the execution of the deobfuscation method are provided before it is called, while the latter can also be the result of other accesses or computations.
Unfortunately, none of the obfuscation techniques that we surveyed in Section~\ref{caseStudy} matches these conditions.

\emph{JMD \cite{jmd2019contra}} re-implements deobfuscation logic of known obfuscators~\cite{zkm2019obfus,allatori2019obfus,dasho2019obfus} to execute it with directly-propagated constants. These constants are extracted from previously identified immediate callers of known deobfuscation methods. After the execution of the deobfuscation logic, the calls to this logic is replaced with the revealed strings.
Unlike our approach, JMD does not consider field accesses or other ways to retrieve the propagated values. It identifies obfuscated strings by searching for a specific loading-instruction (\code{LDC}). Therefore, it misses almost all obfuscated strings, which would be produced by the techniques from Section~\ref{caseStudy} because they are loaded by a different instruction (\code{LDC\_W}). 
Additionally, JMD uses a fixed set of method signatures without considering variations or in-lining of deobfuscation logic.
Finally, the deobfuscation logic uses a constant key, but as shown in Section \ref{methodClassifier}, the key varies with each string usage.
JMD's limitations lead to its poor performance: while successfully processing 94\% of the APKs, only 0.01\% of the strings were deobfuscated. 

\emph{Dex-Oracle \cite{dexoracle2018fenton}} searches for deobfuscation methods and executes them in an emulator. 
It uses fixed method signatures to search the app code. Therefore, it misses variations of methods produced the same obfuscator and in-lined deobfuscation code.  
For instance, only two kinds of signatures for deobfuscation methods are processed.
Whereas, one has only a \code{String} parameter, the other takes three \code{int} parameters. Both signatures return a \code{String}. However, some obfuscators use methods with more than three parameters, which may also have other types than \code{String} or \code{int} and return \code{Object} instead of \code{String}. 
Moreover, it has similar drawbacks as Simplify and JMD -- wherein is it required that the inputs of the deobfuscation method call are instructions that return a constant value.

As \figurename~\ref{fig_accuracy} shows, Dex-Oracle processed all APKs without errors but recovered only 2.5\% of all obfuscated strings.
Its strict assumptions match only very few deobfuscation methods found in the wild, leading to a low recall.
Even those are only a coincidence because the obfuscator, which produced these deobfuscation methods, has various other templates (cf. Section \ref{methodClassifier}) as also shown in Table~\ref{tab:obfuscationTechniques} with \code{cn.pro.sdk}. 

\emph{\sh} was able to process all APKs with a recall of 73.9\%. 
A detailed analysis of the 26.1\% missing cases showed that every obfuscation scheme listed in Table~\ref{tab:obfuscationTechniques} occurred in the false negative set. 
Furthermore, the analysis revealed that either the execution environment, surrounding the sliced method, is too complex to be modeled with our default values (cf. Section~\ref{discussion}) or the classifiers were not able to identify the obfuscated strings (cf. Section~\ref{methodClassifier}).
However, the high recall confirms the effectiveness of our approach, which does not suffer from the various limitations of the state-of-the-art deobfuscators. Unlike our approach, other deobfuscators do not 'automatically' identify obfuscated strings and deobfuscation methods. To use them, one either needs to know the deobfuscation methods beforehand or must run all methods of the app being analyzed. Such a brute-force approach does not scale to large data sets.

\subsection{Findings in the Wild}
\label{findings}
In this section, we use \sh to assess how often string obfuscation is used in the wild and for what purposes. 
Four different sets of APKs are used for our study. 
The first set consists of 100,000 apps from AndroidZoo \cite{hurier2016lack}. The second set consists of the Top 500 most common apps based on AndroidRank~\cite{androidrank2019ranking}.
The third set consists of apps that were available on the Play Store in 2018 and were classified as malicious by at least 10 AV vendors in VirusTotal. 
Finally, the last set consists of 230 Android malware samples from Contagio~\cite{contagio2019dump},
containing current and past malware families.

\subsubsection{Prevalence of Obfuscated Strings in the Wild}
\label{prevalence}
In this section, we measure the prevalence of obfuscated strings in the wild.
Therefore, we apply our approach to 100,000 apps from Section~\ref{caseStudy}.
To avoid false positives, we exclude all constant strings from our findings and count the remaining ones, which we refer to as newly revealed strings.

Next, we calculated the number of APKs containing newly revealed strings. 
Based on our study in Section \ref{caseStudy}, we discovered that only parts of the strings are obfuscated, and some obfuscators hide obfuscated strings in other data structures.

\greybox{Our results invalidate the claims of previous studies~\cite{dong_2018_understanding, mirzaei2018androdet, wang2017changed} that less than 5\% of the apps contain obfuscated strings, because we discovered that 76\% of the 100,000 apps contain obfuscated strings.}

During our analysis, we also measured the proportion of newly revealed strings which were found by the different classifiers. The results indicate that the string classifier detected 28\%, and the method classifier 77\% of the newly revealed strings. 

\greybox{These findings provide empirical evidence that both classifiers are needed because they have only an overlap of 5\% for newly revealed strings. The method classifier identifies most of the newly revealed strings. However, the string classifier provides at least 23\% of newly revealed strings and, thus, it is also necessary to achieve a higher total recall.}

\subsubsection{Categorization of String Obfuscation}
\label{categories}
\begin{figure}[t!]
    \includegraphics[width=\columnwidth]{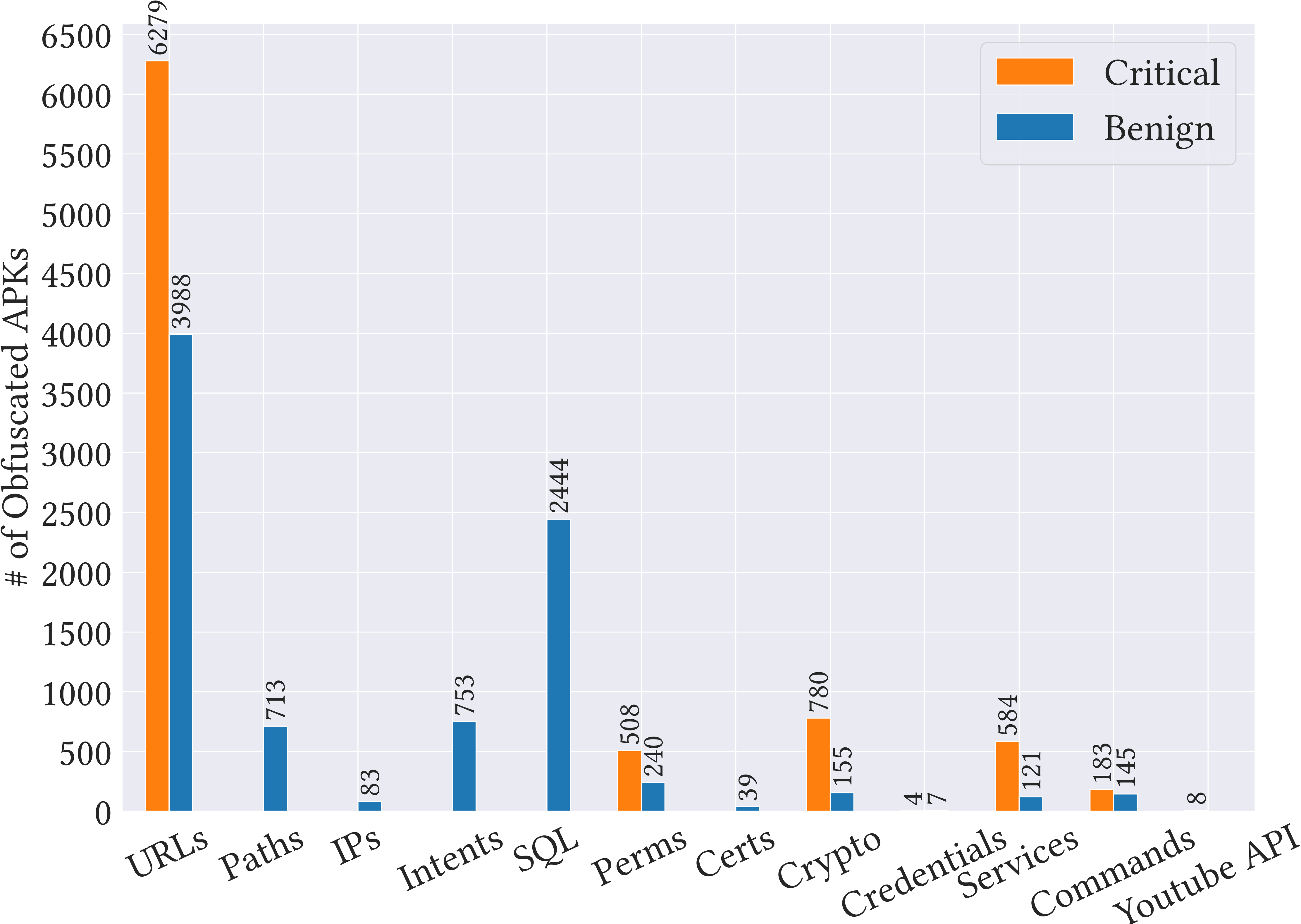}
    \caption{Categories of String Obfuscation in the 100k Apps.}
    \label{fig_100apps}
\end{figure}

To understand the usage of string obfuscation in the wild, we used regular expressions to categorize all deobfuscated strings in different classes.
Altogether, we defined the regular expressions in Table~\ref{tab:regex} for matching URLs, file system paths (Paths), IPs, intents, SQL statements (SQL), permissions (Perms), certificates (Certs), cryptography algorithms (crypto), credentials, system services (Services), commands, and API keys\cite{melibad2019}. 
Our regex for URLs is not limited to the typical HTTP(S) form but also matches any scheme, such as content and file URLs. Furthermore, a pattern for IPs is used to match non-URL related communication.
The third regex matches absolute directories as well as paths pointing to data or executable files.
Intents and permissions are identified via a regex based on Android's standard definitions.
The regular expression for SQL statements matches strings with common keywords for querying and manipulating tables.
Finally, certificates are identified by the Base64 encoded first three characters, which are used as a prefix for certificates.

We applied the regular expressions to all deobfuscated strings in our data sets (discarding apps without newly revealed strings). Afterward, we counted their matches to quantify the prevalence of each kind of usage.

\begin{table}
\caption{Regular Expressions for the Evaluation of Deobfuscated Strings in the Wild.}
\footnotesize
\begin{tabularx}{\columnwidth}{l|l}
\hline
\textbf{Name} & \textbf{Regex}                                                                                 \\ 
\hline
URL           & \textbackslash{}w+://{[}\textasciicircum{}/\textbackslash{}"{]}+.*                             \\
IP            & .*\textbackslash{}b({[}0-9{]}\{1,3\}\textbackslash{}.)\{3\}{[}0-9{]}\{1,3\}\textbackslash{}b.* \\
Paths         & /\textbackslash{}w+{[}\textbackslash{}./{]}.+                                                  \\
Intents       & android.intent\textbackslash{}..*                                                              \\
SQL           & .*(select.*from | update.*set | insert into | \\ 
& delete from |  create table | drop table | \\ 
& truncate table).*  \\
Certificates & MII.+                                                                                          \\
Permissions   & android.permission\textbackslash{}..*\\ 
YouTube API Key~\cite{melibad2019} & AIza{[}0-9A-Za-z\textbackslash{}-\_{]}\{35\}\\
Cryptography algorithms & MD2|MD5|SHA\textbackslash{}-?1|ECB|DES\\
\hline                                   
\end{tabularx}
\label{tab:regex}
\end{table}

\figurename~\ref{fig_100apps} shows a categorization of the resulting deobfuscated strings in the 100,000 apps. The bar chart is divided into critical and benign apps with obfuscated strings. We identified many critical strings that we found by counting the following facts.
First, we identified more HTTP requests than HTTPS for which lead to security issues~\cite{http2019}. 
Second, developers request permissions but are not aware that these permissions are also used via obfuscated strings by ad libraries to access private data. 
Third, insecure cryptography algorithms such as DES, AES with ECB mode, or MD2 are still used in obfuscated strings. 
Fourth, credentials, hidden in obfuscated strings, are sent via HTTP GET method to login to their services. 
Fifth, services, requested in obfuscated strings, provide dangerous accesses (e.g., the location of the device).  
Sixth, rooted phones execute commands, hidden in obfuscated strings, to grant root access.  
Last, YouTube API keys, hidden in obfuscated strings, can be used to consume the developer's API quotas.

\greybox{Using \sh our analysis of the 100,000 apps reveals that critical usages of URLs, piggy-backed permissions, insecure cryptography algorithms, hard-coded credentials, dangerous services, root commands, and API keys are hidden in obfuscated strings.}

\subsubsection{Context Analysis of the Categories}
\begin{figure}[t!]
    \includegraphics[width=\columnwidth]{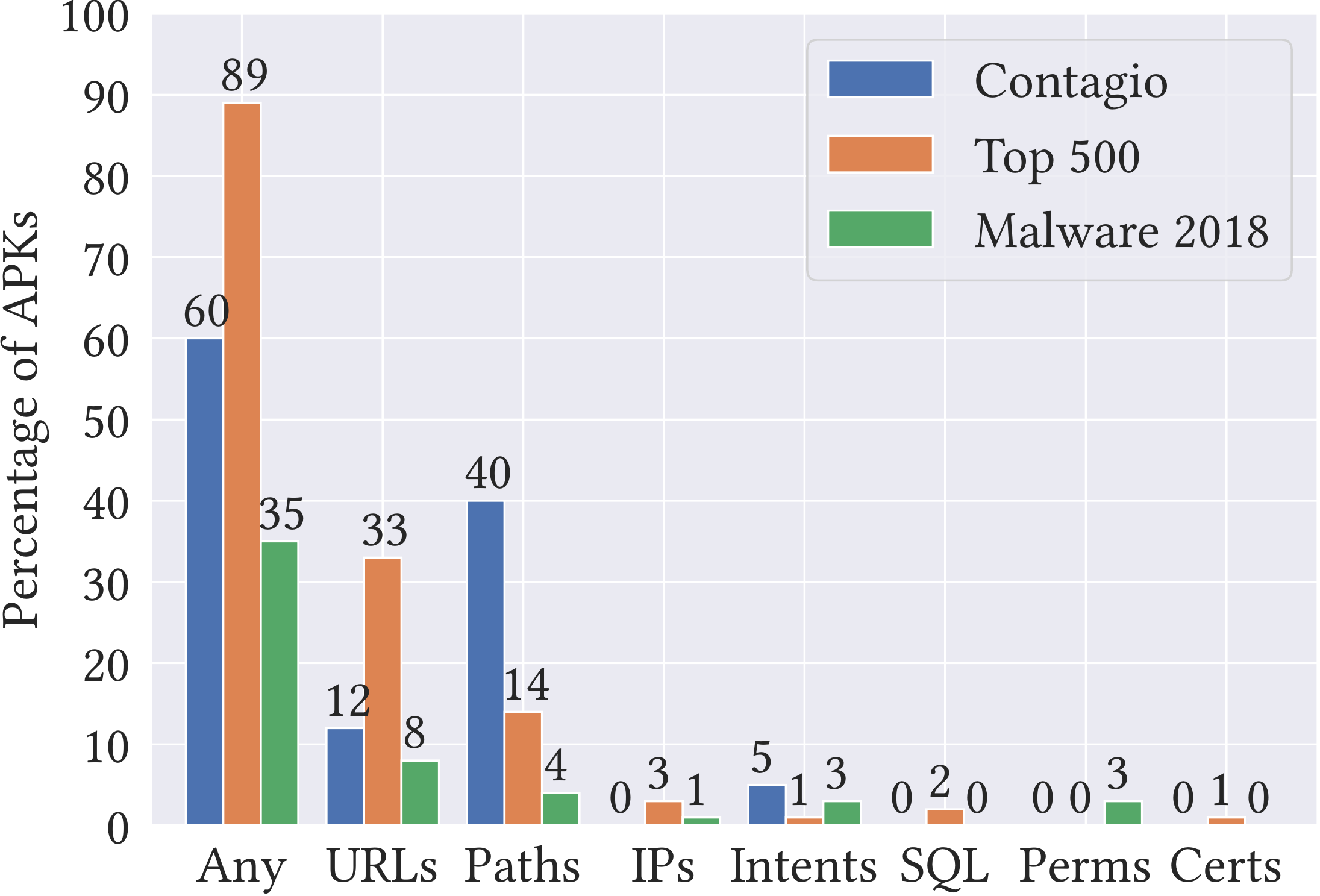}
    \caption{Categories of String Obfuscation in the Wild.}
    \label{fig_regex}
\end{figure}

While the 100,000 apps contain a large variety of statistical findings, we have no insights into apps that belong to the extreme fields in the Android ecosystem. Therefore, we chose three different data sets to get an understanding of these kinds of fields and the context of \sh{}'s findings. These data sets consist of the top 500 most installed apps in the Play store, and two malware sets to analyze current (malware 2018) and past (Contagio) obfuscated malware. 
\figurename~\ref{fig_regex} shows a categorization of the resulting deobfuscated strings. Each bar corresponds to the percentage of APKs from a data set containing at least one deobfuscated string in the given category. Therefore, each category comprises a group of three bars, where each corresponds to one data set.

The first bar shows that 60\% of the Contagio malware obfuscates strings, mostly paths (40\%), URLs (12\%), or intents (5\%).
A detailed analysis revealed absolute paths of commands trying to open a command shell or of further APK or DEX files hidden in the resources of the app containing the malware's actual payload. 
One path was used to establish a connection to a Command \& Control server (AnserverBot~\cite{grace_2012_riskranker}).
Furthermore, we found paths to files on the SD card and to DHCP settings, which are exploited by the DroidKungFu2 malware~\cite{killam2016android}. 
Our regex for URLs matched locations of browser settings that can be used to build a profile of the underlying mobile phone - some URLs linked to services providing the geolocation of the accessing IP address. 
We also found URLs to ad networks that profile the user's phone. The regular expression for intents matched an action that resets the default page of the browser to either show pages of ad networks or to track user's behavior. 
Furthermore, the intent regex discovered an action, which queries the phone number of the mobile phone to reveal the identity of the user. 
Finally, we also found an action that performs phone calls.

In the malware set from 2018, 35\% of APKs use string obfuscation to hide a variety of interactions with the Android operating system. We matched URLs that lead to ad networks, which track the user's interactions and build profiles of users as well as URLs that access the user's calendar and can reveal detailed information of their schedules. The tracking of interactions, in combination with profiling, violates the user's privacy. We also found hidden paths of an APK holding its malicious payload. Moreover, paths to operating-system commands, which access hardware and sensor data to profile a phone, were revealed. Findings regarding intents and permissions indicate that malware uses intents to access functionality to call or send SMS to premium numbers. Additionally, the malware tries to locate or profile a user by accessing personal calendars, accounts, or states of a phone. In comparison to malware from the Contagio set, more recent malware focuses on leakages of private data, causing financial damage to the unknowing user.

\greybox{Current malware in the Play store makes less use of string obfuscation (35\% compared to 60\%) and focuses more on hiding leakages of private data. Without \sh, one would miss information that is essential to detect remote command execution, even causing financial damage to the user, and leakages of private data in at least 35\% of recent malware.}

Surprisingly, string obfuscation is more frequently used in the Top 500 apps than in the 100,000-set of apps (89\% vs. 76\%), even more frequently than in malware. Our evaluation shows that 33\% of the apps use obfuscated URLs. Some of those URLs are used to track users' IDs and IP through an ad network.

These actions directly violate users' privacy. 
A detailed review of the findings showed that all ad libraries contain obfuscated URLs and paths. 
We also analyzed how many apps use string obfuscation only in ad- and third-party libraries\footnote{To this end, we filtered our findings by the list of ad-library package names from Section \ref{caseStudy} and by a list of common libraries~\cite{li2015investigation-tr}.}. This analysis revealed that 63.52\% of all obfuscated strings in the Top 500 data set are contained in ad libraries, an additional 10.64\% are contained in other libraries, and the remaining 25.84\% are in the app itself.

\greybox{String obfuscation is frequently used in all sorts of apps. Ad libraries are responsible for over 63\% of these strings. This result is alarming since neither the user nor the developer of the app is aware of the added functionality. With \sh the developer could check the content of the used ad library and choose an appropriate alternative.}

Two games for children contained obfuscated privacy violations in the Top 500 data set. 
We manually analyzed their code and found that they collect and transmit sensitive information on the user's mobile phone. 
The leaked information includes build, connectivity, debug, runtime, telephony, Android version, and hardware data, which can be used to build a user profile. 
Code related to data collection is hidden in a stealthy package mixed into the integrated Android support library. 
The app additionally checks for the \code{SuperUser.apk}, a package that grants root access to the mobile phone. 
We reported both apps to the app store, since according to AndroidRank~\cite{androidrank2019ranking}, these suspicious apps are installed on at least 20 million devices.
\greybox{Virus scanners do not flag suspicious privacy violations. Since, we uploaded the two apps to VirusTotal~\cite{virustotal2019search}, which showed no findings, besides the usage of dangerous permissions. \sh allows the analyst to search for all kinds of violations.}

%% file: 5_discussion.tex
\section{Discussion}
\label{discussion}
There are a few limitations of 
\sh that will be subject of further consideration in future work. 

Driven by the study of obfuscation schemes, \sh uses intra-procedural slicing to recover automatically obfuscated strings. 
As a result, the slice's execution may fail if it expects values, which differ from our injected defaults. However, this limitation can be addressed by fuzzing the expected values. 
Given a field or parameter, fuzzing guesses their values by their data-dependencies or using symbolic execution to discover possible value ranges. 
If a decryption key is not present in the app code, e.g. because it is downloaded over a network only under very specific circumstances, we cannot deobfuscate strings that are encrypted with it.
Obfuscators can use fields and parameters to perform inter-procedural obfuscation. However, to perform it automatically, they need to identify the call order of the fields and parameters. This call-order is not easily identifiable because of the limitation of current call graph analyses for Android.
Of course, making \sh inter-procedural is an obvious alternative, but coping with potential inter-procedural obfuscation schemes is a trade-off between soundness and performance.

Further challenges for \sh are the dynamic usage of external packages, encrypted classes, and native code. 
While these techniques would evade our approach, the combination of other tools~\cite{dongwoo2015} with \sh can mitigate these challenges.

If an obfuscator adds random dictionary words to a string, it can eventually evade detection by our String classifier because the proportion of content that is classified as non-obfuscated will increase. However, for this technique to be effective, more than half of a given string would need to consist of non-obfuscated words.
During our analysis of obfuscation techniques, we never found more than one dictionary word in obfuscated strings.

Finally, if a new obfuscation technique for strings is used that does not share any commonalities with known techniques, we need to extend the approach with the found technique without training from scratch, as we can train additional REPTrees, and consider a string as obfuscated, as long as at least one REPTree classifies it as such. 
Similarly we can add the SPR representation of a different technique to the list of known obfuscation techniques to adapt the method classification.

%% file: 6_relatedwork.tex
\section{Related work}
\label{relatedWork}

In this section, we discuss four approaches, which could potentially be used for deobfuscation, and studies on the usage of string obfuscation in the wild.

\subsection{Potential Deobfuscation Approaches}
While different slicing approaches \cite{menezes_2017_detecting, hoffmann_slicing_2013, chen2017mass, continella2017obfuscation} exist that could be modified with much effort to deobfuscate strings; others can be used almost directly.
Unfortunately, we could not include the other works~\cite{zhou_2015_harvesting,rasthofer_harvesting_2016,bello2018ares,wong2018tackling} in our empirical evaluation because they were not publicly available. We contacted all authors via e-mail, however, without any responses. Additionally, the re-implementation of their tools was also not possible because some parts cannot be reconstructed from the papers. As a result, we only discuss these approaches in the following based on their descriptions.

\emph{Harvester~\cite{rasthofer_harvesting_2016}}, \emph{TIRO~\cite{wong2018tackling}}, \emph{CredMiner~\cite{zhou_2015_harvesting}}, and \emph{ARES~\cite{bello2018ares}} combine static and dynamic analysis to extract obfuscated runtime values, including obfuscated strings, from Android malware. 
All these approaches execute re-bundled code on an emulator using monkey scripts. This re-bundled code is sliced, starting from a fixed set of starting points.

On the contrary, \sh requires neither re-bundling the app nor an emulator setup that explores all paths with a monkey script. As a result, \sh can analyze Android and Java applications without searching for the correct combination of events to trigger the deobfuscation.
Additionally, our classifiers identify more than a fixed set of starting points.

\subsection{Identifying Obfuscated Apps}
Several approaches have been proposed to identify whether the content of an (Android) app is changed by
an obfuscator \cite{wermke2018large, dong_2018_understanding, mirzaei2018androdet, wang2017changed}. 
While OBFUSCAN~\cite{wermke2018large} only identifies name obfuscations, the other three approaches \cite{dong_2018_understanding, mirzaei2018androdet, wang2017changed} can identify whether the code contains obfuscated strings. Wang et al.~\cite{wang2017changed} even infer the used obfuscator. 
Like our classifiers, all four approaches rely on machine learning techniques to identify whether code is obfuscated or not. However, unlike our approach, they can only detect string obfuscation if all strings in the app are obfuscated. Additionally, they cannot handle obfuscated strings which are represented by byte arrays. We use the token distribution~\cite{glanz2017codematch} with the Spearman's correlation to perform a scalable and lightweight similarity measurement. Other approaches, such as those used in clone detection~\cite{li2017cclearner}, are not suited for obfuscated clones and would require a considerable ground truth for the training of their neuronal networks.

%% file: 7_concl.tex
\section{Conclusion \& Future Work}
\label{conclusion}

This paper shows how and why string obfuscation is used in real-world Android and Java apps.
We presented \sh{}, our approach to identify obfuscated strings and recover their plain text. \sh significantly improves over state-of-the-art deobfuscation tools.
We also presented a large-scale study on the usage of string obfuscation in benign and malicious apps, revealing highly-relevant findings.

We provide empirical evidence that string obfuscation is commonly used across malware, 100,000 apps from Google's Play Store, and various ad libraries. This evidence invalidates statements by previous research, suggesting that string obfuscation is rarely used in practice. 
By undoing string obfuscation, we revealed abundant problematic string usages in the wild: Critical internet accesses, piggy-backed permissions, insecure usage of cryptography algorithms, hard-coded passwords, and available YouTube API keys. 
We have found not only malware concealing hidden commands and communication endpoints, but also spyware-like behavior in two apps in the Top 500 set.
Our studies have shown that libraries account for a significant amount of obfuscated strings in benign apps. Many findings in the ad libraries reveal serious privacy issues.

We have already mentioned several interesting areas for future work in Section~\ref{discussion}. In addition, we will investigate ways to improve \sh{}'s runtime performance by incorporating library detection \cite{wang2018orlis,li2017libd,glanz2017codematch,ma2016libradar,backes2016reliable} and extraction and/or by parallel execution of slices.

%% file: 8_appendix.tex
\section{Appendix}
\label{appendix}
In this section, we describe the runtime performance of \sh. 

\subsection{Runtime Performance}
\label{runtime}
To evaluate the runtime performance of \sh, we measured the average runtime per APK and per 
slice, when running \sh on the Top 500, and the two data sets from Section \ref{findings}.
While the first measure shows how long our approach needs for APKs of different sizes, the second one can be used to approximate the analysis time for a given APK. All performance measures indicate that \sh is fast and ready for practical use.

\begin{figure}[b]
    \includegraphics[width=\columnwidth]{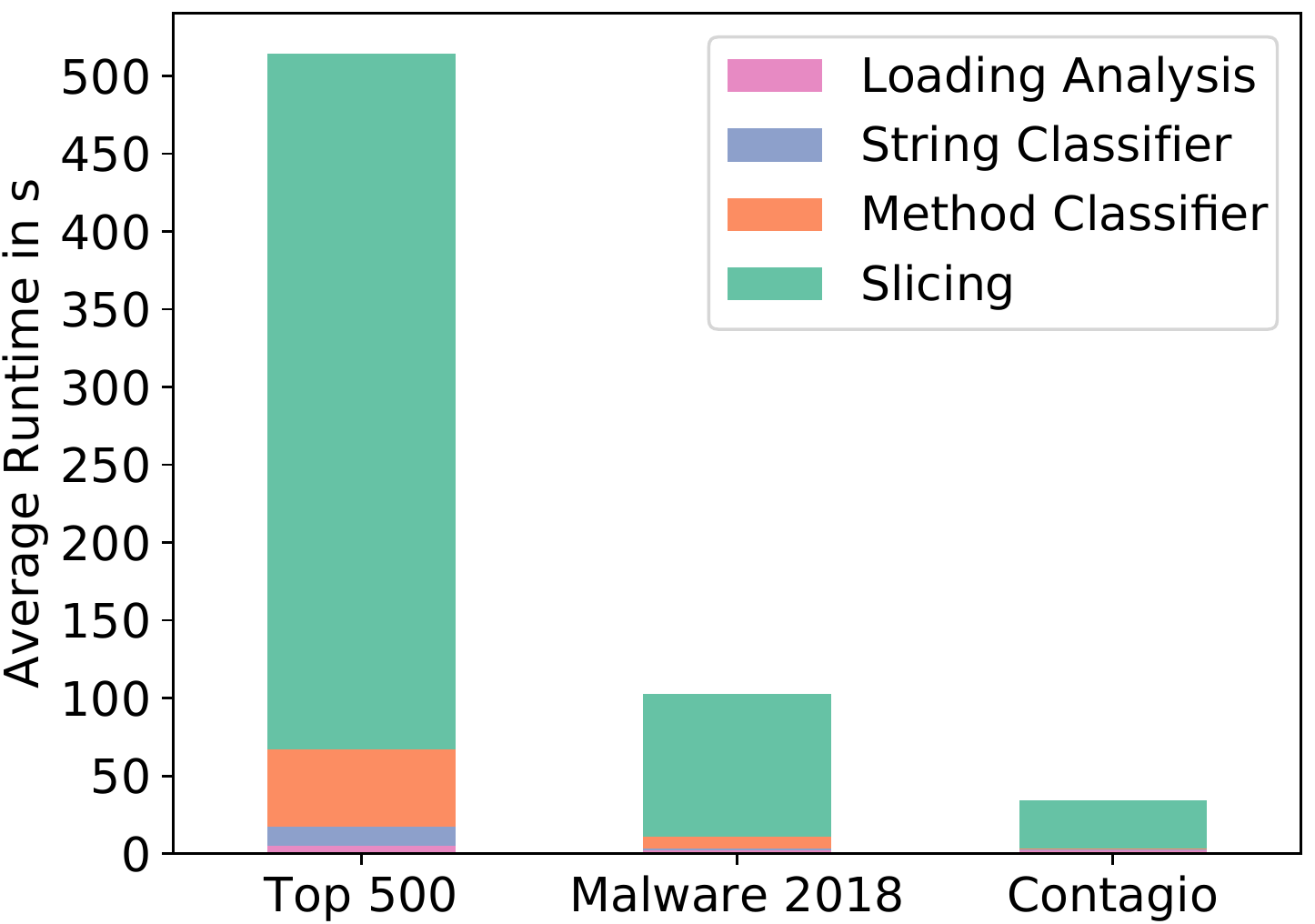}
    \caption{Average Runtime for Top 500 and the Two Malware Data Sets}
    \label{fig_avgtime}
\end{figure}

\figurename~\ref{fig_avgtime} shows the average runtime per APK. 
Thereby, each bar corresponds to one data set and is split into the time needed (a) for loading the analysis, (b) executing the String Classifier, (c) executing the Method Classifier, and (d) building and executing slices. 
One can see, processing the Top 500 data set needs up to 20 times more on average per APK than processing the Contagio data set. The reason for this high discrepancy is a large amount of library code in the APKs of the Top 500 data set. As mentioned in Section \ref{findings}, 74.16\% of the obfuscated strings are found in libraries, and these are up to 14 times larger in code size than APKs from the Contagio data set. The time taken to analyze such apps can be reduced by employing tools that separately analyze the library code and reuse these analysis results. 
Another observation is that across all three data sets, slicing consumes most of the execution time. Hence, improving the performance of the slicing would speed up the entire analysis.

We calculated the mean, median, and also the 95\%-quantile for each slice of all three data sets, and all of them are below 250 ms.
Thus, we conclude that building and executing a single slice takes on average less than 250ms. Given the observation that slicing consumes most of the execution time and also the execution of a single slice takes less than 250 ms, the only improvement to speed up the performance is to parallelize the building and execution of single slices.

%% file: main.bbl

\begin{thebibliography}{65}


\ifx \showCODEN    \undefined \def \showCODEN     #1{\unskip}     \fi
\ifx \showDOI      \undefined \def \showDOI       #1{#1}\fi
\ifx \showISBNx    \undefined \def \showISBNx     #1{\unskip}     \fi
\ifx \showISBNxiii \undefined \def \showISBNxiii  #1{\unskip}     \fi
\ifx \showISSN     \undefined \def \showISSN      #1{\unskip}     \fi
\ifx \showLCCN     \undefined \def \showLCCN      #1{\unskip}     \fi
\ifx \shownote     \undefined \def \shownote      #1{#1}          \fi
\ifx \showarticletitle \undefined \def \showarticletitle #1{#1}   \fi
\ifx \showURL      \undefined \def \showURL       {\relax}        \fi
\providecommand\bibfield[2]{#2}
\providecommand\bibinfo[2]{#2}
\providecommand\natexlab[1]{#1}
\providecommand\showeprint[2][]{arXiv:#2}

\bibitem[\protect\citeauthoryear{Aho~Alfred, Ravi, and
  Ullman~Jeffrey}{Aho~Alfred et~al\mbox{.}}{1986}]%
        {aho1986compilers}
\bibfield{author}{\bibinfo{person}{V Aho~Alfred}, \bibinfo{person}{Sethi Ravi},
  {and} \bibinfo{person}{D Ullman~Jeffrey}.} \bibinfo{year}{1986}\natexlab{}.
\newblock \showarticletitle{Compilers: principles, techniques, and tools}.
\newblock \bibinfo{journal}{\emph{Reading: Addison Wesley Publishing Company}}
  (\bibinfo{year}{1986}).
\newblock


\bibitem[\protect\citeauthoryear{Androidrank}{Androidrank}{5 15}]%
        {androidrank2019ranking}
\bibfield{author}{\bibinfo{person}{Androidrank}.} \bibinfo{year}{Accessed:
  2019-05-15}\natexlab{}.
\newblock
\newblock
\newblock
\shownote{\url{https://www.androidrank.org/}.}


\bibitem[\protect\citeauthoryear{{App Brain's Ad Networks}}{{App Brain's Ad
  Networks}}{5 15}]%
        {ad2019networks}
\bibfield{author}{\bibinfo{person}{{App Brain's Ad Networks}}.}
  \bibinfo{year}{Accessed: 2019-05-15}\natexlab{}.
\newblock
\newblock
\newblock
\shownote{\url{https://www.appbrain.com/stats/libraries/ad-networks}.}


\bibitem[\protect\citeauthoryear{Backes, Bugiel, and Derr}{Backes
  et~al\mbox{.}}{2016}]%
        {backes2016reliable}
\bibfield{author}{\bibinfo{person}{Michael Backes}, \bibinfo{person}{Sven
  Bugiel}, {and} \bibinfo{person}{Erik Derr}.} \bibinfo{year}{2016}\natexlab{}.
\newblock \showarticletitle{Reliable third-party library detection in android
  and its security applications}. In \bibinfo{booktitle}{\emph{Proceedings of
  the 2016 ACM SIGSAC Conference on Computer and Communications Security}}
  \emph{(\bibinfo{series}{CCS'16})}. ACM, \bibinfo{pages}{356--367}.
\newblock


\bibitem[\protect\citeauthoryear{Bello and Pistoia}{Bello and Pistoia}{2018}]%
        {bello2018ares}
\bibfield{author}{\bibinfo{person}{Luciano Bello} {and} \bibinfo{person}{Marco
  Pistoia}.} \bibinfo{year}{2018}\natexlab{}.
\newblock \showarticletitle{ARES: triggering payload of evasive Android
  malware}. In \bibinfo{booktitle}{\emph{2018 IEEE/ACM 5th International
  Conference on Mobile Software Engineering and Systems}}
  \emph{(\bibinfo{series}{MOBILESoft'18})}. IEEE, \bibinfo{pages}{2--12}.
\newblock


\bibitem[\protect\citeauthoryear{Binkley and Gallagher}{Binkley and
  Gallagher}{1996}]%
        {binkley1996program}
\bibfield{author}{\bibinfo{person}{David Binkley} {and}
  \bibinfo{person}{Keith~Brian Gallagher}.} \bibinfo{year}{1996}\natexlab{}.
\newblock \showarticletitle{Program slicing}.
\newblock \bibinfo{journal}{\emph{Advances in Computers}} \bibinfo{volume}{43},
  \bibinfo{number}{1-50} (\bibinfo{year}{1996}), \bibinfo{pages}{1--2}.
\newblock


\bibitem[\protect\citeauthoryear{Chen, You, Lee, Chen, Wang, and Zou}{Chen
  et~al\mbox{.}}{2017}]%
        {chen2017mass}
\bibfield{author}{\bibinfo{person}{Yi Chen}, \bibinfo{person}{Wei You},
  \bibinfo{person}{Yeonjoon Lee}, \bibinfo{person}{Kai Chen},
  \bibinfo{person}{XiaoFeng Wang}, {and} \bibinfo{person}{Wei Zou}.}
  \bibinfo{year}{2017}\natexlab{}.
\newblock \showarticletitle{Mass discovery of android traffic imprints through
  instantiated partial execution}. In \bibinfo{booktitle}{\emph{Proceedings of
  the 2017 ACM SIGSAC Conference on Computer and Communications Security}}
  \emph{(\bibinfo{series}{CCS'17})}. ACM, \bibinfo{pages}{815--828}.
\newblock


\bibitem[\protect\citeauthoryear{class, Android APK~obfuscator, and
  merger}{class et~al\mbox{.}}{2 12}]%
        {shield4j2020}
\bibfield{author}{\bibinfo{person}{Shield4J A~Java class},
  \bibinfo{person}{shrinker Android APK~obfuscator, encrypter}, {and}
  \bibinfo{person}{merger}.} \bibinfo{year}{Accessed: 2020-02-12}\natexlab{}.
\newblock \bibinfo{booktitle}{}.
\newblock
\urldef\tempurl%
\url{https://dzone.com/articles/shield4j-java-class-and}
\showURL{%
\tempurl}


\bibitem[\protect\citeauthoryear{{Contagio Mobile Dump}}{{Contagio Mobile
  Dump}}{5 15}]%
        {contagio2019dump}
\bibfield{author}{\bibinfo{person}{{Contagio Mobile Dump}}.}
  \bibinfo{year}{Accessed: 2019-05-15}\natexlab{}.
\newblock
\newblock
\newblock
\shownote{\url{http://contagiominidump.blogspot.com/}.}


\bibitem[\protect\citeauthoryear{Continella, Fratantonio, Lindorfer, Puccetti,
  Zand, Kruegel, and Vigna}{Continella et~al\mbox{.}}{2017}]%
        {continella2017obfuscation}
\bibfield{author}{\bibinfo{person}{Andrea Continella}, \bibinfo{person}{Yanick
  Fratantonio}, \bibinfo{person}{Martina Lindorfer},
  \bibinfo{person}{Alessandro Puccetti}, \bibinfo{person}{Ali Zand},
  \bibinfo{person}{Christopher Kruegel}, {and} \bibinfo{person}{Giovanni
  Vigna}.} \bibinfo{year}{2017}\natexlab{}.
\newblock \showarticletitle{Obfuscation-Resilient Privacy Leak Detection for
  Mobile Apps Through Differential Analysis.}. In
  \bibinfo{booktitle}{\emph{Proceedings of the 2017 Network and Distributed
  Systems Symposium}} \emph{(\bibinfo{series}{NDSS'17})}.
\newblock


\bibitem[\protect\citeauthoryear{Conventions}{Conventions}{4 26}]%
        {oracle2018naming}
\bibfield{author}{\bibinfo{person}{Oracle~Naming Conventions}.}
  \bibinfo{year}{Accessed: 2019-04-26}\natexlab{}.
\newblock
\newblock
\newblock
\shownote{\url{https://www.oracle.com/technetwork/java/codeconventions-135099.html}.}


\bibitem[\protect\citeauthoryear{{DashO}}{{DashO}}{5 15}]%
        {dasho2019obfus}
\bibfield{author}{\bibinfo{person}{{DashO}}.} \bibinfo{year}{Accessed:
  2019-05-15}\natexlab{}.
\newblock
\newblock
\newblock
\shownote{\url{https://www.preemptive.com/}.}


\bibitem[\protect\citeauthoryear{Demetriou, Merrill, Yang, Zhang, and
  Gunter}{Demetriou et~al\mbox{.}}{2016}]%
        {demetriou2016free}
\bibfield{author}{\bibinfo{person}{Soteris Demetriou}, \bibinfo{person}{Whitney
  Merrill}, \bibinfo{person}{Wei Yang}, \bibinfo{person}{Aston Zhang}, {and}
  \bibinfo{person}{Carl~A Gunter}.} \bibinfo{year}{2016}\natexlab{}.
\newblock \showarticletitle{Free for All! Assessing User Data Exposure to
  Advertising Libraries on Android.}. In \bibinfo{booktitle}{\emph{Proceedings
  of the 2016 Annual Network and Distributed System Security Symposium}}
  \emph{(\bibinfo{series}{NDSS'16})}.
\newblock


\bibitem[\protect\citeauthoryear{{Dex Oracle}}{{Dex Oracle}}{5 15}]%
        {dexoracle2018fenton}
\bibfield{author}{\bibinfo{person}{{Dex Oracle}}.} \bibinfo{year}{Accessed:
  2019-05-15}\natexlab{}.
\newblock
\newblock
\newblock
\shownote{\url{https://github.com/CalebFenton/dex-oracle}.}


\bibitem[\protect\citeauthoryear{{Dex2Jar Decrypt Strings}}{{Dex2Jar Decrypt
  Strings}}{5 15}]%
        {dex2jar2019decrypt}
\bibfield{author}{\bibinfo{person}{{Dex2Jar Decrypt Strings}}.}
  \bibinfo{year}{Accessed: 2019-05-15}\natexlab{}.
\newblock
\newblock
\newblock
\shownote{\url{https://sourceforge.net/p/dex2jar/wiki/DecryptStrings/}.}


\bibitem[\protect\citeauthoryear{{DexGuard}}{{DexGuard}}{0 23}]%
        {dexguard2015guardsquare}
\bibfield{author}{\bibinfo{person}{{DexGuard}}.} \bibinfo{year}{Accessed:
  2017-10-23}\natexlab{}.
\newblock
\newblock
\newblock
\shownote{\url{https://www.guardsquare.com/en/dexguard}.}


\bibitem[\protect\citeauthoryear{{Dong, Shuaike and Li, Menghao and Diao,
  Wenrui and Liu, Xiangyu and Liu, Jian and Li, Zhou and Xu, Fenghao and Chen,
  Kai and Wang, Xiaofeng and Zhang, Kehuan}}{{Dong, Shuaike and Li, Menghao and
  Diao, Wenrui and Liu, Xiangyu and Liu, Jian and Li, Zhou and Xu, Fenghao and
  Chen, Kai and Wang, Xiaofeng and Zhang, Kehuan}}{2018}]%
        {dong_2018_understanding}
\bibfield{author}{\bibinfo{person}{{Dong, Shuaike and Li, Menghao and Diao,
  Wenrui and Liu, Xiangyu and Liu, Jian and Li, Zhou and Xu, Fenghao and Chen,
  Kai and Wang, Xiaofeng and Zhang, Kehuan}}.} \bibinfo{year}{2018}\natexlab{}.
\newblock \showarticletitle{{Understanding Android Obfuscation Techniques: A
  Large-Scale Investigation in the Wild}}.
\newblock \bibinfo{journal}{\emph{Springer}} (\bibinfo{year}{2018}),
  \bibinfo{pages}{172--192}.
\newblock


\bibitem[\protect\citeauthoryear{Eichberg and Hermann}{Eichberg and
  Hermann}{2014}]%
        {Eichberg_Hermann_2014}
\bibfield{author}{\bibinfo{person}{Michael Eichberg} {and} \bibinfo{person}{Ben
  Hermann}.} \bibinfo{year}{2014}\natexlab{}.
\newblock \showarticletitle{A Software Product Line for Static Analyses: The
  OPAL Framework}. In \bibinfo{booktitle}{\emph{Proceedings of the 3rd ACM
  SIGPLAN International Workshop on the State of the Art in Java Program
  Analysis}} \emph{(\bibinfo{series}{SOAP '14})}. \bibinfo{publisher}{ACM},
  \bibinfo{pages}{1--6}.
\newblock


\bibitem[\protect\citeauthoryear{Enslen, Hill, Pollock, and
  Vijay-Shanker}{Enslen et~al\mbox{.}}{2009}]%
        {enslen2009mining}
\bibfield{author}{\bibinfo{person}{Eric Enslen}, \bibinfo{person}{Emily Hill},
  \bibinfo{person}{Lori Pollock}, {and} \bibinfo{person}{K Vijay-Shanker}.}
  \bibinfo{year}{2009}\natexlab{}.
\newblock \showarticletitle{Mining source code to automatically split
  identifiers for software analysis}.
\newblock \bibinfo{journal}{\emph{IEEE Computer Society}}
  (\bibinfo{year}{2009}), \bibinfo{pages}{71--80}.
\newblock


\bibitem[\protect\citeauthoryear{{F-Droid}}{{F-Droid}}{5 15}]%
        {fdroid2019store}
\bibfield{author}{\bibinfo{person}{{F-Droid}}.} \bibinfo{year}{Accessed:
  2019-05-15}\natexlab{}.
\newblock
\newblock
\newblock
\shownote{\url{https://f-droid.org/}.}


\bibitem[\protect\citeauthoryear{Fratantonio, Bianchi, Robertson, Kirda,
  Kruegel, and Vigna}{Fratantonio et~al\mbox{.}}{2016}]%
        {fratantonio2016triggerscope}
\bibfield{author}{\bibinfo{person}{Yanick Fratantonio},
  \bibinfo{person}{Antonio Bianchi}, \bibinfo{person}{William Robertson},
  \bibinfo{person}{Engin Kirda}, \bibinfo{person}{Christopher Kruegel}, {and}
  \bibinfo{person}{Giovanni Vigna}.} \bibinfo{year}{2016}\natexlab{}.
\newblock \showarticletitle{Triggerscope: Towards detecting logic bombs in
  android applications}. In \bibinfo{booktitle}{\emph{2016 IEEE Symposium on
  Security and Privacy}} \emph{(\bibinfo{series}{SP'16})}. IEEE,
  \bibinfo{pages}{377--396}.
\newblock


\bibitem[\protect\citeauthoryear{Glanz, Amann, Eichberg, Reif, Hermann, Lerch,
  and Mezini}{Glanz et~al\mbox{.}}{2017}]%
        {glanz2017codematch}
\bibfield{author}{\bibinfo{person}{Leonid Glanz}, \bibinfo{person}{Sven Amann},
  \bibinfo{person}{Michael Eichberg}, \bibinfo{person}{Michael Reif},
  \bibinfo{person}{Ben Hermann}, \bibinfo{person}{Johannes Lerch}, {and}
  \bibinfo{person}{Mira Mezini}.} \bibinfo{year}{2017}\natexlab{}.
\newblock \showarticletitle{CodeMatch: obfuscation won't conceal your
  repackaged app}. In \bibinfo{booktitle}{\emph{Proceedings of the 2017 11th
  Joint Meeting on Foundations of Software Engineering}}. ACM,
  \bibinfo{pages}{638--648}.
\newblock


\bibitem[\protect\citeauthoryear{Grace, Zhou, Zhang, Zou, and Jiang}{Grace
  et~al\mbox{.}}{2012}]%
        {grace_2012_riskranker}
\bibfield{author}{\bibinfo{person}{Michael Grace}, \bibinfo{person}{Yajin
  Zhou}, \bibinfo{person}{Qiang Zhang}, \bibinfo{person}{Shihong Zou}, {and}
  \bibinfo{person}{Xuxian Jiang}.} \bibinfo{year}{2012}\natexlab{}.
\newblock \showarticletitle{Riskranker: scalable and accurate zero-day android
  malware detection}. In \bibinfo{booktitle}{\emph{Proceedings of the 10th
  international conference on Mobile systems, applications, and services}}.
  ACM, \bibinfo{pages}{281--294}.
\newblock


\bibitem[\protect\citeauthoryear{Hoffmann, Ussath, Holz, and
  Spreitzenbarth}{Hoffmann et~al\mbox{.}}{2013}]%
        {hoffmann_slicing_2013}
\bibfield{author}{\bibinfo{person}{Johannes Hoffmann}, \bibinfo{person}{Martin
  Ussath}, \bibinfo{person}{Thorsten Holz}, {and} \bibinfo{person}{Michael
  Spreitzenbarth}.} \bibinfo{year}{2013}\natexlab{}.
\newblock \showarticletitle{Slicing {Droids}: {Program} {Slicing} for {Smali}
  {Code}}. In \bibinfo{booktitle}{\emph{Proceedings of the 28th {Annual} {ACM}
  {Symposium} on {Applied} {Computing}}} \emph{(\bibinfo{series}{{SAC} '13})}.
  \bibinfo{publisher}{ACM}, \bibinfo{address}{New York, NY, USA},
  \bibinfo{pages}{1844--1851}.
\newblock
\showISBNx{978-1-4503-1656-9}


\bibitem[\protect\citeauthoryear{Hurier, Allix, Bissyand{\'e}, Klein, and
  Le~Traon}{Hurier et~al\mbox{.}}{2016}]%
        {hurier2016lack}
\bibfield{author}{\bibinfo{person}{M{\'e}d{\'e}ric Hurier},
  \bibinfo{person}{Kevin Allix}, \bibinfo{person}{Tegawend{\'e}~F
  Bissyand{\'e}}, \bibinfo{person}{Jacques Klein}, {and} \bibinfo{person}{Yves
  Le~Traon}.} \bibinfo{year}{2016}\natexlab{}.
\newblock \showarticletitle{On the lack of consensus in anti-virus decisions:
  Metrics and insights on building ground truths of android malware}. In
  \bibinfo{booktitle}{\emph{International Conference on Detection of Intrusions
  and Malware, and Vulnerability Assessment}}
  \emph{(\bibinfo{series}{DIMVA'16})}. Springer, \bibinfo{pages}{142--162}.
\newblock


\bibitem[\protect\citeauthoryear{{Java bytecode analysis/deobfuscation
  tool}}{{Java bytecode analysis/deobfuscation tool}}{5 15}]%
        {jmd2019contra}
\bibfield{author}{\bibinfo{person}{{Java bytecode analysis/deobfuscation
  tool}}.} \bibinfo{year}{Accessed: 2019-05-15}\natexlab{}.
\newblock
\newblock
\newblock
\shownote{\url{https://github.com/contra/JMD}.}


\bibitem[\protect\citeauthoryear{Kahn}{Kahn}{1996}]%
        {kahn1996codebreakers}
\bibfield{author}{\bibinfo{person}{David Kahn}.}
  \bibinfo{year}{1996}\natexlab{}.
\newblock \bibinfo{booktitle}{\emph{The Codebreakers: The comprehensive history
  of secret communication from ancient times to the internet}}.
\newblock \bibinfo{publisher}{Simon and Schuster}.
\newblock


\bibitem[\protect\citeauthoryear{Killam, Cook, and Stakhanova}{Killam
  et~al\mbox{.}}{2016}]%
        {killam2016android}
\bibfield{author}{\bibinfo{person}{Richard Killam}, \bibinfo{person}{Paul
  Cook}, {and} \bibinfo{person}{Natalia Stakhanova}.}
  \bibinfo{year}{2016}\natexlab{}.
\newblock \showarticletitle{Android malware classification through analysis of
  string literals}.
\newblock \bibinfo{journal}{\emph{Text Analytics for Cybersecurity and Online
  Safety (TA-COS)}} (\bibinfo{year}{2016}).
\newblock


\bibitem[\protect\citeauthoryear{Kim, Kwak, and Ryou}{Kim
  et~al\mbox{.}}{2015}]%
        {dongwoo2015}
\bibfield{author}{\bibinfo{person}{Dongwoo Kim}, \bibinfo{person}{Jin Kwak},
  {and} \bibinfo{person}{Jaecheol Ryou}.} \bibinfo{year}{2015}\natexlab{}.
\newblock \showarticletitle{DWroidDump: Executable Code Extraction from Android
  Applications for Malware Analysis}.
\newblock \bibinfo{journal}{\emph{International Journal of Distributed Sensor
  Networks}} \bibinfo{volume}{11}, \bibinfo{number}{9} (\bibinfo{year}{2015}),
  \bibinfo{pages}{379682}.
\newblock
\urldef\tempurl%
\url{https://doi.org/10.1155/2015/379682}
\showDOI{\tempurl}


\bibitem[\protect\citeauthoryear{Li, Bissyand\'e, Klein, and Le~Traon}{Li
  et~al\mbox{.}}{2015}]%
        {li2015investigation-tr}
\bibfield{author}{\bibinfo{person}{Li Li}, \bibinfo{person}{Tegawend\'e~F.
  Bissyand\'e}, \bibinfo{person}{Jacques Klein}, {and} \bibinfo{person}{Yves
  Le~Traon}.} \bibinfo{year}{2015}\natexlab{}.
\newblock \showarticletitle{{An Investigation into the Use of Common Libraries
  in Android Apps}}. In \bibinfo{booktitle}{\emph{Technique Report}}.
\newblock


\bibitem[\protect\citeauthoryear{Li, Feng, Zhuang, Meng, and Ryder}{Li
  et~al\mbox{.}}{2017a}]%
        {li2017cclearner}
\bibfield{author}{\bibinfo{person}{Liuqing Li}, \bibinfo{person}{He Feng},
  \bibinfo{person}{Wenjie Zhuang}, \bibinfo{person}{Na Meng}, {and}
  \bibinfo{person}{Barbara Ryder}.} \bibinfo{year}{2017}\natexlab{a}.
\newblock \showarticletitle{Cclearner: A deep learning-based clone detection
  approach}. In \bibinfo{booktitle}{\emph{2017 IEEE International Conference on
  Software Maintenance and Evolution}} \emph{(\bibinfo{series}{ICSME'17})}.
  IEEE, \bibinfo{pages}{249--260}.
\newblock


\bibitem[\protect\citeauthoryear{Li, Wang, Wang, Wang, Wu, Liu, Xue, and
  Huo}{Li et~al\mbox{.}}{2017b}]%
        {li2017libd}
\bibfield{author}{\bibinfo{person}{Menghao Li}, \bibinfo{person}{Wei Wang},
  \bibinfo{person}{Pei Wang}, \bibinfo{person}{Shuai Wang},
  \bibinfo{person}{Dinghao Wu}, \bibinfo{person}{Jian Liu},
  \bibinfo{person}{Rui Xue}, {and} \bibinfo{person}{Wei Huo}.}
  \bibinfo{year}{2017}\natexlab{b}.
\newblock \showarticletitle{LibD: scalable and precise third-party library
  detection in android markets}. In \bibinfo{booktitle}{\emph{2017 IEEE/ACM
  39th International Conference on Software Engineering}}
  \emph{(\bibinfo{series}{ICSE'17})}. IEEE, \bibinfo{pages}{335--346}.
\newblock


\bibitem[\protect\citeauthoryear{Lyda and Hamrock}{Lyda and Hamrock}{2007}]%
        {lyda2007using}
\bibfield{author}{\bibinfo{person}{Robert Lyda} {and} \bibinfo{person}{James
  Hamrock}.} \bibinfo{year}{2007}\natexlab{}.
\newblock \showarticletitle{Using entropy analysis to find encrypted and packed
  malware}.
\newblock \bibinfo{journal}{\emph{IEEE Security \& Privacy}}
  \bibinfo{volume}{5}, \bibinfo{number}{2} (\bibinfo{year}{2007}),
  \bibinfo{pages}{40--45}.
\newblock


\bibitem[\protect\citeauthoryear{Ma, Wang, Guo, and Chen}{Ma
  et~al\mbox{.}}{2016}]%
        {ma2016libradar}
\bibfield{author}{\bibinfo{person}{Ziang Ma}, \bibinfo{person}{Haoyu Wang},
  \bibinfo{person}{Yao Guo}, {and} \bibinfo{person}{Xiangqun Chen}.}
  \bibinfo{year}{2016}\natexlab{}.
\newblock \showarticletitle{LibRadar: fast and accurate detection of
  third-party libraries in Android apps}. In
  \bibinfo{booktitle}{\emph{Proceedings of the 38th International Conference on
  Software Engineering}} \emph{(\bibinfo{series}{ICSE'16})}. ACM,
  \bibinfo{pages}{653--656}.
\newblock


\bibitem[\protect\citeauthoryear{Mariconti, Onwuzurike, Andriotis,
  De~Cristofaro, Ross, and Stringhini}{Mariconti et~al\mbox{.}}{2016}]%
        {mariconti2016mamadroid}
\bibfield{author}{\bibinfo{person}{Enrico Mariconti}, \bibinfo{person}{Lucky
  Onwuzurike}, \bibinfo{person}{Panagiotis Andriotis},
  \bibinfo{person}{Emiliano De~Cristofaro}, \bibinfo{person}{Gordon Ross},
  {and} \bibinfo{person}{Gianluca Stringhini}.}
  \bibinfo{year}{2016}\natexlab{}.
\newblock \showarticletitle{Mamadroid: Detecting android malware by building
  markov chains of behavioral models}.
\newblock \bibinfo{journal}{\emph{arXiv preprint arXiv:1612.04433}}
  (\bibinfo{year}{2016}).
\newblock


\bibitem[\protect\citeauthoryear{Meli, McNiece, and Reaves}{Meli
  et~al\mbox{.}}{2019}]%
        {melibad2019}
\bibfield{author}{\bibinfo{person}{Michael Meli}, \bibinfo{person}{Matthew~R
  McNiece}, {and} \bibinfo{person}{Bradley Reaves}.}
  \bibinfo{year}{2019}\natexlab{}.
\newblock \showarticletitle{How Bad Can It Git? Characterizing Secret Leakage
  in Public GitHub Repositories.}. In \bibinfo{booktitle}{\emph{NDSS}}.
\newblock


\bibitem[\protect\citeauthoryear{Menezes and Wism{\"u}ller}{Menezes and
  Wism{\"u}ller}{2017}]%
        {menezes_2017_detecting}
\bibfield{author}{\bibinfo{person}{Luis Menezes} {and} \bibinfo{person}{Roland
  Wism{\"u}ller}.} \bibinfo{year}{2017}\natexlab{}.
\newblock \showarticletitle{Detecting information leaks in Android applications
  using a hybrid approach with program slicing, instrumentation and tagging}.
  In \bibinfo{booktitle}{\emph{Security Technology (ICCST)}}. IEEE,
  \bibinfo{pages}{1--6}.
\newblock


\bibitem[\protect\citeauthoryear{{Mirzaei, O and de Fuentes, JM and Tapiador, J
  and Gonzalez-Manzano, L}}{{Mirzaei, O and de Fuentes, JM and Tapiador, J and
  Gonzalez-Manzano, L}}{2018}]%
        {mirzaei2018androdet}
\bibfield{author}{\bibinfo{person}{{Mirzaei, O and de Fuentes, JM and Tapiador,
  J and Gonzalez-Manzano, L}}.} \bibinfo{year}{{2018}}\natexlab{}.
\newblock \showarticletitle{{AndrODet: An adaptive android obfuscation
  detector}}.
\newblock \bibinfo{journal}{\emph{{Future Generation Computer Systems}}}
  (\bibinfo{year}{{2018}}).
\newblock


\bibitem[\protect\citeauthoryear{Myers and Sirois}{Myers and Sirois}{2004}]%
        {myers2004spearman}
\bibfield{author}{\bibinfo{person}{Leann Myers} {and} \bibinfo{person}{Maria~J
  Sirois}.} \bibinfo{year}{2004}\natexlab{}.
\newblock \showarticletitle{Spearman correlation coefficients, differences
  between}.
\newblock \bibinfo{journal}{\emph{Encyclopedia of statistical sciences}}
  \bibinfo{volume}{12} (\bibinfo{year}{2004}).
\newblock


\bibitem[\protect\citeauthoryear{Nan, Yang, Wang, Zhang, Zhu, and Yang}{Nan
  et~al\mbox{.}}{2018}]%
        {nan2018finding}
\bibfield{author}{\bibinfo{person}{Yuhong Nan}, \bibinfo{person}{Zhemin Yang},
  \bibinfo{person}{Xiaofeng Wang}, \bibinfo{person}{Yuan Zhang},
  \bibinfo{person}{Donglai Zhu}, {and} \bibinfo{person}{Min Yang}.}
  \bibinfo{year}{2018}\natexlab{}.
\newblock \showarticletitle{Finding clues for your secrets: Semantics-driven,
  learning-based privacy discovery in mobile apps}. In
  \bibinfo{booktitle}{\emph{Proceedings of the 2018 Annual Network and
  Distributed System Security Symposium}} \emph{(\bibinfo{series}{NDSS'18})}.
\newblock


\bibitem[\protect\citeauthoryear{Obfuscator}{Obfuscator}{5 15}]%
        {allatori2019obfus}
\bibfield{author}{\bibinfo{person}{Allatori~Java Obfuscator}.}
  \bibinfo{year}{Accessed: 2019-05-15}\natexlab{}.
\newblock
\newblock
\newblock
\shownote{\url{http://www.allatori.com/}.}


\bibitem[\protect\citeauthoryear{Pan, Wang, Duan, Wang, and Yin}{Pan
  et~al\mbox{.}}{2017}]%
        {pan2017dark}
\bibfield{author}{\bibinfo{person}{Xiaorui Pan}, \bibinfo{person}{Xueqiang
  Wang}, \bibinfo{person}{Yue Duan}, \bibinfo{person}{XiaoFeng Wang}, {and}
  \bibinfo{person}{Heng Yin}.} \bibinfo{year}{2017}\natexlab{}.
\newblock \showarticletitle{Dark Hazard: Learning-based, Large-Scale Discovery
  of Hidden Sensitive Operations in Android Apps.}. In
  \bibinfo{booktitle}{\emph{Proceedings of the 2016 Annual Network and
  Distributed System Security Symposium}} \emph{(\bibinfo{series}{NDSS'17})}.
\newblock


\bibitem[\protect\citeauthoryear{{Practical Cryptography}}{{Practical
  Cryptography}}{5 15}]%
        {cryptanalysis2019lyons}
\bibfield{author}{\bibinfo{person}{{Practical Cryptography}}.}
  \bibinfo{year}{Accessed: 2019-05-15}\natexlab{}.
\newblock
\newblock
\newblock
\shownote{\url{http://practicalcryptography.com/cryptanalysis/}.}


\bibitem[\protect\citeauthoryear{provides minimal obfuscation. DexGuard applies
  multiple layers~of encryption and obfuscation.}{provides minimal obfuscation.
  DexGuard applies multiple layers~of encryption and obfuscation.}{2017}]%
        {provsdex2020}
\bibfield{author}{\bibinfo{person}{ProGuard provides minimal obfuscation.
  DexGuard applies multiple layers~of encryption} {and}
  \bibinfo{person}{obfuscation.}} \bibinfo{year}{2017}\natexlab{}.
\newblock \bibinfo{booktitle}{}.
\newblock
\urldef\tempurl%
\url{https://www.guardsquare.com/en/blog/dexguard-vs-proguard}
\showURL{%
\tempurl}
\newblock
\shownote{\url{https://www.guardsquare.com/en/blog/dexguard-vs-proguard},
  2020-02-18.}


\bibitem[\protect\citeauthoryear{Quinlan}{Quinlan}{1986}]%
        {quinlan1986induction}
\bibfield{author}{\bibinfo{person}{J.~Ross Quinlan}.}
  \bibinfo{year}{1986}\natexlab{}.
\newblock \showarticletitle{Induction of decision trees}.
\newblock \bibinfo{journal}{\emph{Machine learning}} \bibinfo{volume}{1},
  \bibinfo{number}{1} (\bibinfo{year}{1986}), \bibinfo{pages}{81--106}.
\newblock


\bibitem[\protect\citeauthoryear{Rasthofer, Arzt, Miltenberger, and
  Bodden}{Rasthofer et~al\mbox{.}}{2016}]%
        {rasthofer_harvesting_2016}
\bibfield{author}{\bibinfo{person}{Siegfried Rasthofer},
  \bibinfo{person}{Steven Arzt}, \bibinfo{person}{Marc Miltenberger}, {and}
  \bibinfo{person}{Eric Bodden}.} \bibinfo{year}{2016}\natexlab{}.
\newblock \showarticletitle{Harvesting {Runtime} {Values} in {Android}
  {Applications} {That} {Feature} {Anti}-{Analysis} {Techniques}}. In
  \bibinfo{booktitle}{\emph{NDSS}}.
\newblock


\bibitem[\protect\citeauthoryear{Razaghpanah, Nithyanand, Vallina-Rodriguez,
  Sundaresan, Allman, and Gill}{Razaghpanah et~al\mbox{.}}{2018}]%
        {razaghpanah2018apps}
\bibfield{author}{\bibinfo{person}{Abbas Razaghpanah}, \bibinfo{person}{Rishab
  Nithyanand}, \bibinfo{person}{Narseo Vallina-Rodriguez},
  \bibinfo{person}{Srikanth Sundaresan}, \bibinfo{person}{Mark Allman}, {and}
  \bibinfo{person}{Christian Kreibich~Phillipa Gill}.}
  \bibinfo{year}{2018}\natexlab{}.
\newblock \showarticletitle{Apps, trackers, privacy, and regulators}. In
  \bibinfo{booktitle}{\emph{25th Annual Network and Distributed System Security
  Symposium, NDSS}}, Vol.~\bibinfo{volume}{2018}.
\newblock


\bibitem[\protect\citeauthoryear{Schrittwieser, Katzenbeisser, Kinder,
  Merzdovnik, and Weippl}{Schrittwieser et~al\mbox{.}}{2016}]%
        {schrittwieser2016protecting}
\bibfield{author}{\bibinfo{person}{Sebastian Schrittwieser},
  \bibinfo{person}{Stefan Katzenbeisser}, \bibinfo{person}{Johannes Kinder},
  \bibinfo{person}{Georg Merzdovnik}, {and} \bibinfo{person}{Edgar Weippl}.}
  \bibinfo{year}{2016}\natexlab{}.
\newblock \showarticletitle{Protecting software through obfuscation: Can it
  keep pace with progress in code analysis?}
\newblock \bibinfo{journal}{\emph{ACM Computing Surveys (CSUR)}}
  \bibinfo{volume}{49}, \bibinfo{number}{1} (\bibinfo{year}{2016}),
  \bibinfo{pages}{4}.
\newblock


\bibitem[\protect\citeauthoryear{{Simplify}}{{Simplify}}{5 15}]%
        {simplify2019fenton}
\bibfield{author}{\bibinfo{person}{{Simplify}}.} \bibinfo{year}{Accessed:
  2019-05-15}\natexlab{}.
\newblock
\newblock
\newblock
\shownote{\url{https://github.com/CalebFenton/simplify}.}


\bibitem[\protect\citeauthoryear{Son, Kim, and Shmatikov}{Son
  et~al\mbox{.}}{2016}]%
        {son2016mobile}
\bibfield{author}{\bibinfo{person}{Sooel Son}, \bibinfo{person}{Daehyeok Kim},
  {and} \bibinfo{person}{Vitaly Shmatikov}.} \bibinfo{year}{2016}\natexlab{}.
\newblock \showarticletitle{What Mobile Ads Know About Mobile Users.}. In
  \bibinfo{booktitle}{\emph{Proceedings of the 2016 Network and Distributed
  Systems Symposium}} \emph{(\bibinfo{series}{NDSS'16})}.
\newblock


\bibitem[\protect\citeauthoryear{Stevens, Gibler, Crussell, Erickson, and
  Chen}{Stevens et~al\mbox{.}}{2012}]%
        {stevens2012investigating}
\bibfield{author}{\bibinfo{person}{Ryan Stevens}, \bibinfo{person}{Clint
  Gibler}, \bibinfo{person}{Jon Crussell}, \bibinfo{person}{Jeremy Erickson},
  {and} \bibinfo{person}{Hao Chen}.} \bibinfo{year}{2012}\natexlab{}.
\newblock \showarticletitle{Investigating user privacy in android ad
  libraries}. In \bibinfo{booktitle}{\emph{Workshop on Mobile Security
  Technologies}} \emph{(\bibinfo{series}{MoST'12})}, Vol.~\bibinfo{volume}{10}.
\newblock


\bibitem[\protect\citeauthoryear{{Stringer Java Obfuscator}}{{Stringer Java
  Obfuscator}}{5 15}]%
        {stringer2014obfus}
\bibfield{author}{\bibinfo{person}{{Stringer Java Obfuscator}}.}
  \bibinfo{year}{Accessed: 2019-05-15}\natexlab{}.
\newblock
\newblock
\newblock
\shownote{\url{https://jfxstore.com/}.}


\bibitem[\protect\citeauthoryear{users with TLS by default~in Android~P}{users
  with TLS by default~in Android~P}{1 22}]%
        {http2019}
\bibfield{author}{\bibinfo{person}{Protecting users with TLS by default~in
  Android~P}.} \bibinfo{year}{Accessed: 2019-11-22}\natexlab{}.
\newblock \bibinfo{booktitle}{}.
\newblock
\urldef\tempurl%
\url{https://android-developers.googleblog.com/2018/04/protecting-users-with-tls-by-default-in.html}
\showURL{%
\tempurl}


\bibitem[\protect\citeauthoryear{Vidas and Christin}{Vidas and
  Christin}{2014}]%
        {vidas2014evading}
\bibfield{author}{\bibinfo{person}{Timothy Vidas} {and}
  \bibinfo{person}{Nicolas Christin}.} \bibinfo{year}{2014}\natexlab{}.
\newblock \showarticletitle{Evading android runtime analysis via sandbox
  detection}. In \bibinfo{booktitle}{\emph{Proceedings of the 9th ACM symposium
  on Information, computer and communications security}}. ACM,
  \bibinfo{pages}{447--458}.
\newblock


\bibitem[\protect\citeauthoryear{{VirusTotal}}{{VirusTotal}}{5 15}]%
        {virustotal2019search}
\bibfield{author}{\bibinfo{person}{{VirusTotal}}.} \bibinfo{year}{Accessed:
  2019-05-15}\natexlab{}.
\newblock
\newblock
\newblock
\shownote{\url{https://www.virustotal.com/}.}


\bibitem[\protect\citeauthoryear{Wang and Rountev}{Wang and Rountev}{2017}]%
        {wang2017changed}
\bibfield{author}{\bibinfo{person}{Yan Wang} {and} \bibinfo{person}{Atanas
  Rountev}.} \bibinfo{year}{2017}\natexlab{}.
\newblock \showarticletitle{Who changed you?: obfuscator identification for
  Android}. In \bibinfo{booktitle}{\emph{Proceedings of the 4th International
  Conference on Mobile Software Engineering and Systems}}.
  \bibinfo{pages}{154--164}.
\newblock


\bibitem[\protect\citeauthoryear{Wang, Wu, Zhang, and Rountev}{Wang
  et~al\mbox{.}}{2018}]%
        {wang2018orlis}
\bibfield{author}{\bibinfo{person}{Yan Wang}, \bibinfo{person}{Haowei Wu},
  \bibinfo{person}{Hailong Zhang}, {and} \bibinfo{person}{Atanas Rountev}.}
  \bibinfo{year}{2018}\natexlab{}.
\newblock \showarticletitle{Orlis: Obfuscation-resilient library detection for
  Android}. In \bibinfo{booktitle}{\emph{2018 IEEE/ACM 5th International
  Conference on Mobile Software Engineering and Systems}}
  \emph{(\bibinfo{series}{MOBILESoft'18})}. IEEE, \bibinfo{pages}{13--23}.
\newblock


\bibitem[\protect\citeauthoryear{Wermke, Huaman, Acar, Reaves, Traynor, and
  Fahl}{Wermke et~al\mbox{.}}{2018}]%
        {wermke2018large}
\bibfield{author}{\bibinfo{person}{Dominik Wermke}, \bibinfo{person}{Nicolas
  Huaman}, \bibinfo{person}{Yasemin Acar}, \bibinfo{person}{Bradley Reaves},
  \bibinfo{person}{Patrick Traynor}, {and} \bibinfo{person}{Sascha Fahl}.}
  \bibinfo{year}{2018}\natexlab{}.
\newblock \showarticletitle{A large scale investigation of obfuscation use in
  google play}. In \bibinfo{booktitle}{\emph{Proceedings of the 34th Annual
  Computer Security Applications Conference}}
  \emph{(\bibinfo{series}{ACSAC'18})}. ACM, \bibinfo{pages}{222--235}.
\newblock


\bibitem[\protect\citeauthoryear{WhatsApp}{WhatsApp}{5 15}]%
        {whatsapp2018messenger}
\bibfield{author}{\bibinfo{person}{WhatsApp}.} \bibinfo{year}{Accessed:
  2019-05-15}\natexlab{}.
\newblock
\newblock
\newblock
\shownote{\url{https://play.google.com/store/apps/details?id=com.whatsapp}.}


\bibitem[\protect\citeauthoryear{Wong and Lie}{Wong and Lie}{2016}]%
        {wong2016intellidroid}
\bibfield{author}{\bibinfo{person}{Michelle~Y Wong} {and}
  \bibinfo{person}{David Lie}.} \bibinfo{year}{2016}\natexlab{}.
\newblock \showarticletitle{IntelliDroid: A Targeted Input Generator for the
  Dynamic Analysis of Android Malware.}. In
  \bibinfo{booktitle}{\emph{Proceedings of the 2016 Annual Network and
  Distributed System Security Symposium}} \emph{(\bibinfo{series}{NDSS'16})},
  Vol.~\bibinfo{volume}{16}. \bibinfo{pages}{21--24}.
\newblock


\bibitem[\protect\citeauthoryear{Wong and Lie}{Wong and Lie}{2018}]%
        {wong2018tackling}
\bibfield{author}{\bibinfo{person}{Michelle~Y Wong} {and}
  \bibinfo{person}{David Lie}.} \bibinfo{year}{2018}\natexlab{}.
\newblock \showarticletitle{Tackling runtime-based obfuscation in Android with
  {TIRO}}. In \bibinfo{booktitle}{\emph{27th {USENIX} Security Symposium}}
  \emph{(\bibinfo{series}{USENIX Security'18})}. \bibinfo{pages}{1247--1262}.
\newblock


\bibitem[\protect\citeauthoryear{{Zelix KlassMaster}}{{Zelix KlassMaster}}{5
  15}]%
        {zkm2019obfus}
\bibfield{author}{\bibinfo{person}{{Zelix KlassMaster}}.}
  \bibinfo{year}{Accessed: 2019-05-15}\natexlab{}.
\newblock
\newblock
\newblock
\shownote{\url{http://www.zelix.com/}.}


\bibitem[\protect\citeauthoryear{Zhao, Zuo, Pellegrino, and Zhiqiang}{Zhao
  et~al\mbox{.}}{2019}]%
        {zhao2019geo}
\bibfield{author}{\bibinfo{person}{Qingchuan Zhao}, \bibinfo{person}{Chaoshun
  Zuo}, \bibinfo{person}{Giancarlo Pellegrino}, {and} \bibinfo{person}{Li
  Zhiqiang}.} \bibinfo{year}{2019}\natexlab{}.
\newblock \showarticletitle{Geo-locating Drivers: A Study of Sensitive Data
  Leakage in Ride-Hailing Services.}. In \bibinfo{booktitle}{\emph{Proceedings
  of the 2019 Annual Network and Distributed System Security Symposium}}
  \emph{(\bibinfo{series}{NDSS'19})}.
\newblock


\bibitem[\protect\citeauthoryear{Zhou, Wu, Wang, and Jiang}{Zhou
  et~al\mbox{.}}{2015}]%
        {zhou_2015_harvesting}
\bibfield{author}{\bibinfo{person}{Yajin Zhou}, \bibinfo{person}{Lei Wu},
  \bibinfo{person}{Zhi Wang}, {and} \bibinfo{person}{Xuxian Jiang}.}
  \bibinfo{year}{2015}\natexlab{}.
\newblock \showarticletitle{Harvesting developer credentials in android apps}.
  In \bibinfo{booktitle}{\emph{Proceedings of the 8th ACM Conference on
  Security \& Privacy in Wireless and Mobile Networks}}. ACM,
  \bibinfo{pages}{23}.
\newblock


\bibitem[\protect\citeauthoryear{Zuo, Lin, and Zhang}{Zuo
  et~al\mbox{.}}{2019}]%
        {zuo2019does}
\bibfield{author}{\bibinfo{person}{Chaoshun Zuo}, \bibinfo{person}{Zhiqiang
  Lin}, {and} \bibinfo{person}{Yinqian Zhang}.}
  \bibinfo{year}{2019}\natexlab{}.
\newblock \showarticletitle{Why does your data leak? uncovering the data
  leakage in cloud from mobile apps}. In \bibinfo{booktitle}{\emph{IEEE
  Symposium on Security and Privacy}} \emph{(\bibinfo{series}{SP'19})}.
\newblock


\end{thebibliography}
